\documentclass[11pt,reqno]{JHEP3}
\usepackage[dvips]{graphicx}
\usepackage{epsfig}

\usepackage{amsmath,amssymb}
\allowdisplaybreaks[1]

\def\be{\begin{equation}}
\def\ee{\end{equation}}
\def\Zop{\mathbb{Z}}
\def\N{{\cal N}}

\def\la{{\lambda}}
\def\be{{\beta}}

\def\i{{\rm i}}

\title{Twisted brane charges for non-simply 
connected groups}
\author{Matthias R.\ Gaberdiel\\ 
Institut f\"ur Theoretische Physik, 
ETH Z\"urich\\
8093 Z\"urich, Switzerland\\
E-mail: \email{gaberdiel@itp.phys.ethz.ch}}
\author{Terry Gannon\\
Department of Mathematical Sciences, University of
Alberta \\
Edmonton, Alberta, Canada, T6G 2G1 \\
E-mail: \email{tgannon@math.ualberta.ca}}
\abstract{The charges of the twisted branes for strings on the group
manifold SU$(n)/\Zop_d$ are determined. To this end we derive explicit
(and remarkably simple) formulae for the relevant NIM-rep
coefficients. The charge groups of the twisted and untwisted 
branes are compared and found to agree for the cases we consider.}
\preprint{}
\begin{document}

\section{Introduction and Background}

The charges of D-branes in string theory are believed to be
characterised in terms of K-theory
\cite{Minasian:1997mm,Witten:1998cd,Moore:2003vf}. For example, for
strings that propagate on a group manifold $G$, the charge group is
conjectured to be the twisted K-theory $^{k+h^\vee}K (G)$ 
\cite{Kapustin:1999di,Bouwknegt:2000qt}, where the twist involves 
the Wess-Zumino form of the underlying Wess-Zumino-Witten (WZW) model at 
level $k$.

For all simple, simply connected Lie groups $G$, the twisted K-theory
has been computed in \cite{Braun:2003rd} (see also \cite{FHT,D})   
to be
\begin{equation}\label{Kgroup}
^{k+h^\vee}K (G)\ =\ 
\underbrace{\mathbb{Z}_{M(\bar{\mathfrak{g}},k)}\oplus \dotsb \oplus 
\mathbb{Z}_{M(\bar{\mathfrak{g}},k)}}_{2^{{\rm rk} 
(\bar{\mathfrak{g}})-1}}\equiv 2^{{\rm rk}(\bar{\mathfrak{g}})-1}\cdot
\Zop_{M(\bar{\mathfrak{g}},k)}\  ,
\end{equation}
where $M(\bar{\mathfrak{g}},k)$ is the integer 
\begin{equation}
M(\bar{\mathfrak{g}},k)\  =\  \frac{k+h^\vee}{{\rm gcd}(k+h^\vee,L)} \ . 
\end{equation}
Here $h^\vee$ is the dual Coxeter number of the finite dimensional
Lie algebra $\bar{\mathfrak{g}}$ of rank 
${\rm rk}(\bar{\mathfrak{g}})$, and $L$ only depends on   
$\bar{\mathfrak{g}}$ (but not on $k$). In fact, except for the case of
$C_n$ that will not concern us in this paper, $L$ is 
\begin{equation}
L \ =\  {\rm lcm} \{1,2,\ldots,h-1\} \ ,
\end{equation} 
where $h$ is the Coxeter number of $\bar{\mathfrak{g}}$. For
$\bar{\mathfrak{g}}=A_n$ this formula was derived in 
\cite{Fredenhagen:2000ei,Bouwknegt:2002bq} (see also \cite{MMS}), 
while the formulae in the other cases were checked numerically up to
very high levels in \cite{Bouwknegt:2002bq}. For the classical Lie
algebras and $G_2$ an alternative expression for $M$ was also found in
\cite{D}. If $G$ is a non-simply connected group manifold much less
is known; so far only the case of SO(3) = SU(2) $/\Zop_2$ has been
worked out in detail \cite{Braun:2004qg}.
\smallskip

The results of these K-theory analyses should be compared with what
can be determined directly in terms of the underlying conformal field
theory. The idea behind this approach is that brane configurations
that are connected by RG flows should carry the same charge. For the  
branes $x\in \mathcal{B}_{k}^{\omega}$ of an arbitrary WZW model that
preserve the full affine  symmetry algebra $\mathfrak{g}$ up to some
automorphism $\omega$, this constraint implies in particular, that
their charges $q(x)$ satisfy \cite{Fredenhagen:2000ei} 
\begin{equation}\label{chargerelation}
\dim (\lambda)\  q(x) \ =\ \sum_{y\in \mathcal{B}_{k}^{\omega }} 
\mathcal{N}_{\lambda x}{}^{y}\ q(y)  \ .
\end{equation}
Here $\lambda\in \mathcal{P}_{k}^{+} (\bar{\mathfrak{g}})$ is an
arbitrary  highest-weight representation of the affine Lie algebra
$\mathfrak{g}$ at level $k$, $\dim(\lambda)$ is the Weyl-dimension of
the corresponding representation of the horizontal subalgebra
$\bar{\mathfrak{g}}$, and $\mathcal{N}_{\lambda a}{}^{b}$ are the
NIM-rep coefficients appearing in the Cardy analysis (for an
introduction to these matters see \cite{BPPZ,gannon}).

For the case of the simply connected group manifold, 
the corresponding WZW model is the charge-conjugation modular
invariant; by analogy to the ADE classification for the case of SU(2)
we shall in the following refer to it as the ${\cal A}$-modular
invariant. The untwisted branes (that correspond to the trivial
automorphism $\omega =\text{id}$) can be labelled by integrable highest
weights of $\mathfrak{g}$, 
$\mathcal{B}_{k}^{\text{id}}\cong
\mathcal{P}_{k}^{+}(\bar{\mathfrak{g}})$, and the NIM-rep 
${\cal N}({\cal A})$ agrees with the fusion rules. In this case, the  
constraints~\eqref{chargerelation} were evaluated in
\cite{Fredenhagen:2000ei,Bouwknegt:2002bq}. The charges are given (up
to rescalings) by the Weyl-dimensions of the corresponding
representations, $q(\lambda)=\dim (\lambda)$, and the charge is
conserved only modulo $M(\bar{\mathfrak{g}},k)$. Thus, the untwisted
branes account for one summand 
$\mathbb{Z}_{M(\bar{\mathfrak{g}},k)}$ of the K-group~\eqref{Kgroup}.   

For non-trivial outer automorphisms, a similar analysis was carried
through in \cite{Gaberdiel:2003kv,GV,Fredenhagen:2005cj}. Here, the
D-branes are parametrised by $\omega$-twisted highest weight
representations $a$ of $\mathfrak{g}_{k}$
\cite{Birke:1999ik,Fuchs:2000vg,Gaberdiel:2002qa}, 
and the NIM-rep coefficients are given by twisted fusion rules
\cite{Gaberdiel:2002qa}. In fact, the NIM-rep is the same as the one
describing the untwisted branes of the ${\cal A}^\omega$-modular
invariant that is obtained from the charge conjugation 
${\cal A}$-modular invariant by applying the automorphism $\omega$ to
the left-movers say; for the case of SU$(n)$ that shall concern us in
the following, $\omega$ is charge conjugation and we shall thus denote
the corresponding NIM-rep by ${\cal N}({\cal A}^*)$. 

The twisted representations can be identified with representations of
the invariant subalgebra $\bar{\mathfrak{g}}^{\omega}$ consisting of 
$\omega$-invariant elements of $\bar{\mathfrak{g}}$, and we can view
$\mathcal{B}_{k}^{\omega}$ as a subset of
$\mathcal{P}_{k'}^{+} (\bar{\mathfrak{g}}^{\omega })$, where 
$k'=k+ h^\vee(\bar{\mathfrak{g}}) -h^\vee(\bar{\mathfrak{g}}^{\omega})$. 
It was found in \cite{Gaberdiel:2003kv} that the charge $q(a)$ 
of $a\in \mathcal{B}_{k}^{\omega}$  
is again (up to rescalings) given by the Weyl dimension\footnote{A
related proposal was made in~\cite{Alekseev:2002rj} 
based on an analysis for large level.} of the
representation of $\bar{\mathfrak{g}}^{\omega}$, $q(a)=\dim (a)$, and
that the charge identities are only satisfied modulo 
$M(\bar{\mathfrak{g}},k)$. Thus each such class of twisted D-branes
accounts for another summand  $\mathbb{Z}_{M(\bar{\mathfrak{g}},k)}$
of the charge group.  Since the number of 
automorphisms does not grow with the rank, these 
constructions do not in general account for all the charges of
(\ref{Kgroup}); for the case of the $A_n$ series, a proposal for the
D-branes that may carry the remaining charges was made in 
\cite{Gaberdiel:2004hs,Gaberdiel:2004za} (see also
\cite{MMS}). 
\medskip

In this paper we shall study the D-branes charges for string theory
on the non-simply connected group manifolds SU$(n)/\Zop_d$, where $d$ is
a factor of $n$. The corresponding modular invariants
are known \cite{FGK1,FGK2}. By analogy to the 
SU(2) case we shall call them the ${\cal D}$-modular invariants.  
For quotient groups of SU$(n)$ there are then two classes of branes
that preserve the full affine symmetry up to an automorphism. First
there are the untwisted branes for which the automorphism is trivial;
the corresponding NIM-rep will be denoted by 
${\cal N}({\cal D})$. Some aspects of this  NIM-rep were already
studied in \cite{Gaberdiel:2004yn} where also the charge group was
partially determined. As was observed there, the analysis depends
crucially on whether $n(n+1)/d$ is even or odd. In particular, it was
found that if $n(n+1)/d$ is odd, the charge group is surprisingly
small; this was called the {\em pathological case} in
\cite{Gaberdiel:2004yn}. This pathological behaviour may be related to
the fact that in these cases there is a second modular invariant that
one can consider (in which the fermionic degrees are treated
differently); at least for the case of SO(3) the charge group of this
second theory has again the expected size  \cite{Fredenhagen:2004xp}.
 
In this paper we shall only consider the non-pathological case, 
{\it i.e.}\ we shall assume that $n(n+1)/d$ is even. For this class of 
theories we shall be able to give a complete description of the
NIM-rep ${\cal N}({\cal D})$. This will also allow us to determine the 
${\cal D}$ charge group in more detail; this will be described in
section~4.

The main new results of this paper however concern the analysis of the
twisted D-branes, {\it i.e.}\ the branes that preserve the affine
symmetry up to the outer automorphism that corresponds to charge
conjugation. Since these branes are naturally in one-to-one
correspondence with the untwisted branes of the modular invariant
${\cal D}^*$ (that can be obtained from ${\cal D}$ by charge
conjugation) we shall denote the relevant NIM-rep by 
${\cal N}({\cal D}^*)$. As we shall see, we can give a remarkably
simple formula for this NIM-rep in all non-pathological cases. In many
cases we can furthermore determine the resulting charge groups in
detail. This will be explained in sections~2 and 3. 
\medskip

For the case of the simply connected groups it was shown in 
\cite{Gaberdiel:2003kv,GV,Fredenhagen:2005cj} that the charge groups
of the untwisted (${\cal A}$) and twisted (${\cal A}^*$) branes
coincide. One may wonder whether this is true in general. For the
non-pathological cases we study in this paper this also seems to be
the case: whenever we can determine both the 
${\cal D}$- and ${\cal D}^*$-charge groups explicitly, they
agree. This agreement is fairly non-trivial since the calculations
that are involved for the two cases are at least superficially very
different. One may therefore expect that there is a more conceptual
explanation of this correspondence; some ideas in this direction are
described in section~5. We also note there that this agreement only
seems to hold in the non-pathological case; we have found an explicit
counterexample, namely SU$(4)/\Zop_4$ at level $k=4$ (which is
pathological), for which the two charge groups disagree.
\bigskip

The paper is organised as follows. In the following subsection we
introduce some more notation. We construct the ${\cal D}^*$ NIM-rep
and determine its charge group for the case when $n$ is odd (which is
always non-pathological) in section~2. The non-pathological cases with
$n$ even are dealt with in section~3. In section~4 we give a more
detailed description of the ${\cal D}$ NIM-rep in the non-pathological
cases and calculate the charge group explicitly (at least for certain
classes of examples). A possible relation between the charge groups of
${\cal D}$ and ${\cal D}^*$ is explained in section~5, and section~6
contains some conclusions. We have included a number of appendices in
which some technical proofs are given.

\subsection{Some background material and notation}

Before we can begin with the detailed discussion we need to introduce
some notation. Recall that the integrable highest weights 
$\la\in P_+^{k}({\rm su}(n))$ consist of all $n$-tuples
$\la=(\la_0;\la_1,\ldots,\la_{n-1})$ of non-negative integers  
$\la_i$, where $\sum_{j=0}^{n-1}\la_j=k$. We will often drop the
redundant $0$'th Dynkin label $\la_0$. 
We shall also often write $\widehat{{\rm su}}(n)_k$ for `affine
su$(n)$ at level $k$'. Charge-conjugation $C$ and the 
generator $J$ of simple currents is given by
\begin{eqnarray}
\la^* & =&\,C\la=(\la_0;\la_{n-1},\la_{n-2},\ldots,\la_1) \\
J\la & =&\,(\la_{n-1};\la_0,\la_1,\ldots,\la_{n-2})\ .
\end{eqnarray}
Their effect on the $S$-matrix elements is 
$S_{C\la,\mu}=\overline{S_{\la,\mu}}$ and 
\begin{equation}
S_{J^j\la,\mu}=\exp[2\pi {\rm i} j\,t(\mu)/n]\,S_{\la,\mu}\ , 
\end{equation}
where $t(\mu)=\sum_{j=1}^{n-1}j\mu_j$ is the $n$-ality of $\mu$.
Note that $C(J^j\la)=J^{-j}C\la$,  and $t(C\la)= -t(\la)$ 
(mod $n$). 

\noindent Simple currents give rise to symmetries and gradings of
fusion rule coefficients 
\begin{eqnarray}
&& N_{J^i\lambda,J^j\mu}{}^{J^{i+j}\nu} = \,N_{\la\mu}{}^\nu
\label{1.1a} \\
&& N_{\la\mu}{}^\nu\ne 0\ \Longrightarrow\  t(\la)+t(\mu)=t(\nu)
\quad ({\rm mod}\ n)\ . \label{1.1b}
\end{eqnarray}

Strings moving on the simply connected covering group
SU$(n)$ \cite{FGK1,FGK2} are described by the WZW model with modular 
invariant 
\begin{equation}
M_{\lambda\mu} = \delta_{\lambda\mu} \ .
\end{equation}
This modular invariant is sometimes referred to as the 
${\cal A}$-modular invariant.

The centre of SU$(n)$ is $\Zop_n$. For any factor $d$ of $n$, we
therefore have a quotient group SU$(n)/\Zop_d$. The corresponding
\cite{FGK1,FGK2} modular invariant is \cite{SY}
\begin{equation}\label{modular}
M[d']_{\la\mu}=
\sum_{j=1}^{d}\delta_{{d}}
\left(t(\la)+\frac{d'\, j\, k'}{2}\right)\,
\delta^{\mu\, J^{j d'}\la}\ , 
\end{equation}
where $d'=n/d$, and $k'=k+n$ if $k$ and $n$ are odd, and $k'=k$
otherwise. In order for this to define a modular invariant partition
function we need to have that $n(n+1)k/d$ is even. 
If this is the case, we call it the ${\cal D}$-modular invariant.  

The simple current $J^{d'}$ that will play an important
role in the following, has order $d$. We call $\varphi\in P_+^{k}
({\rm su}(n))$ a
{\it fixed point of order $m$} (with respect to $J^{d'}$) if
the $J^{d'}$-orbit $\{J^{jd'}\varphi\}$ has cardinality $d/m$ -- in
other words,  $m$
divides $d$, and $d/m$ is the smallest positive integer for which  
$J^{d'(d/m)}\varphi = J^{n/m}\varphi=\varphi$. 
 Write $o(\varphi)$ for the order $m$ of
$\varphi$. Note that any solution $\varphi\in P_+^k({\rm su}(n))$ to
$J^{n/m}\varphi=\varphi$ looks 
like $\varphi=(\bar{\varphi},\ldots,\bar{\varphi})$ 
($m$ copies of $\bar{\varphi}$),  where 
$\bar{\varphi} =(\varphi_0;\ldots,\varphi_{n/m-1})\in 
P_+^{k/m}({\rm su}(n/m))$.
For a given $n,k,d$, $J^{d'}$ will have fixed points of order $m$,
when and only when $m$ divides gcd$(d,k)$.
If a fixed point $\varphi$ has order $m$, and $\Delta$ divides $m$,
then by $\varphi^\Delta$ we mean the 
$\widehat{\rm su}(n/\Delta)_{k/\Delta}$ weight 
obtained by retaining only the first $n/\Delta$ components of
$\varphi$.  
\smallskip

A NIM-rep ${\cal N}$ for $\widehat{{\rm su}}(n)_k$ can be uniquely
specified in two different ways: either by giving the matrices 
${\cal N}_{\Lambda_m}=({\cal N}_{\Lambda_m\ x}{}^y)$ for all
fundamental weights $\Lambda_m=(0,\ldots,1,\ldots,0)$; or by
specifying the matrix $\psi=(\psi_{x\mu})$ which simultaneously
diagonalises all matrices 
${\cal N}_\lambda=({\cal N}_{\lambda x}{}^y)$ in the sense that  
\begin{equation}\label{nimpsi}
{\cal N}_{\lambda x}{}^y=\sum_{\mu}\psi_{x\mu}\,
\frac{S_{\lambda \mu}}{S_{0 \mu}}\,
\psi_{y \mu}^*\ .
\end{equation}
Here, $\lambda\in P_+^k({\rm su}(n))$, $x,y$ are boundary labels, and
$\mu$ are the exponents for the corresponding NIM-rep, {\it i.e.}\ the
weights that label the possible Ishibashi states (with multiplicity). 
Alternatively, these are just the exponents of the corresponding
modular invariant ${\cal Z}$, {\it i.e.}\ the weights appearing with
multiplicity ${\cal Z}_{\mu\mu}$. 

Given a NIM-rep ${\cal N}$, we can determine the corresponding charge
group following \cite{Fredenhagen:2000ei}. In particular, a 
{\it charge assignment} consists of integers $q(x)$ 
(one for each boundary label) and $M$ such that (\ref{chargerelation})
is satisfied mod $M$. For $\widehat{{\rm su}}(n)_k$ we have 
$h^\vee = h=n$, and thus 
\begin{equation} \label{M_A}
M \equiv M({\rm su}(n),k) = 
\frac{n+k}{{\rm gcd}(n+k,{\rm lcm}(1,2,\ldots,n-1))} \ .
\end{equation}
Furthermore, one knows on general grounds that any charge assignment
is always proportional to one defined modulo $M$.
The sum $q(x)\equiv q'(x)+q''(x)$ of arbitrary charge assignments
$q'(x),q''(x)$ (defined mod $M$) is likewise one defined mod $M$.
The charge assignments thus form an additive group called the
{\it charge group}; for $\widehat{{\rm su}}(n)_k$ it takes the form
\begin{equation}\label{chgp}
\Zop_{M_1}\oplus\cdots\oplus\Zop_{M_t}\ ,
\end{equation}
where $\Zop_m$ denotes the additive group $\Zop/m\Zop$, and where each
$M_i$ divides $M$. The decomposition (\ref{chgp}) becomes unique if in
addition we require each $M_i$ to divide $M_{i-1}$. Both ${\cal A}$
and ${\cal A}^*$ have $t=1$, but both ${\cal D}$ and ${\cal D}^*$ can
have $t>1$. 
\smallskip

After these preliminary remarks we can now study the 
${\cal D}^*$ NIM-rep and its associated charge group.

\section{The ${\cal D}^*$ case with $n$ odd} 

We begin by analysing the charges for ${\cal D}^*$ for $n$ odd. In
section~2.1 we construct the relevant NIM-rep explicitly. In
section~2.2 we then determine its charge group. Given the simplicity
of our formula for the NIM-rep, we are able to do this in closed
form.

\subsection{The NIM-rep}

As before we write $d'=n/d$. Since $n$ is odd, we may write
$n=2m+1$. The NIM-rep for ${\cal A}^*$ on $\widehat{{\rm su}}(n)_k$ was
given in section~4.1 of \cite{Gaberdiel:2002qa}. Recall that the
exponents of the charge-conjugate (${\cal A}^*$) modular invariant for
su($n$) consist of all $\mu\in P_+^{k}({\rm su}(n))$ with  
$C\mu=\mu$, all with multiplicity $1$. The boundary states
$\widehat{a}$ are parametrised by the level $k$ integrable highest
weights of the twisted affine algebra $A_{n-1}^{(2)}$, {\it i.e.}\ all 
$(m+1)$-tuples $(a_0;a_1,\ldots,a_m)$ of non-negative integers where
$k=a_0+2a_1+2a_2+ \cdots+2a_m$. The $\psi$-matrix, diagonalising the
NIM-rep, is the modular $\widehat{S}$-matrix of $A_{n-1}^{(2)}$ (see
eq.~(4.2) in \cite{Gaberdiel:2002qa}). The NIM-rep coefficients are
$C$-twisted fusion coefficients of $A_{n-1,k}$, and can be expressed
in terms of ordinary fusions of both $B_{m}^{(1)}$ level $k+2$  and
$C_{m}^{(1)}$ level $(k-1)/2$
\cite{Gaberdiel:2002qa,Gaberdiel:2003kv}. We will write this NIM-rep
in the form 
\begin{equation}\label{Asni}
\la.\widehat{a}=\sum_{\widehat{b}}{\cal N}
({\cal A}^*)_{\la \widehat{a}}{}^{\widehat{b}}\;\; \widehat{b}\ .
\end{equation} 

For the ${\cal D}^*$ modular invariant ({\it i.e.}\ the charge conjugate
of (\ref{modular})) one finds with a little work that the exponents
form the multi-set $\bigcup_{j=1}^d\bigcup_{\mu=C\mu}J^{jd'}\mu$. That
is, $\mu\in P_+^{k}({\rm su}(n))$ will have multiplicity $0$ unless 
$C(J^{jd'}\mu)=J^{jd'}\mu$ for some $j$, in which case its
multiplicity will equal its order as a $J^{d'}$-fixed point. We should
stress that this simple result assumes that $n$ is odd.

We can thus unambiguously specify the exponents of $M[d']^*$ by pairs 
$(\mu,i)$, where $\mu=C\mu$ and $1\le i\le d$. Likewise, the 
boundary states for $M[d']^*$ are given by all pairs  
$(\widehat{a},j)$, $0\le j<d$. The formula for the ${\cal D}^*$
NIM-rep for any $n,d$ (provided only that $n$ is odd), is given by  
the remarkably simple formula 
\begin{equation}\label{D*odd}
{\cal D}^* : \qquad \boxed{
\la.(\widehat{a},j)=(\la.\widehat{a},j+t(\la)) }
\end{equation}
where the first product is defined by (\ref{Asni}) and 
$j+t(\la)$ is to be taken mod $d$. The $\psi$-matrix for
$M[d']^*$ is  
\begin{equation}
\psi({\cal D}^*)_{(\widehat{a},j),(\mu,i)}
=\frac{1}{\sqrt{d}}e^{2\pi{\rm i} \,ij/d}
\psi({\cal A}^*)_{\widehat{a},\mu}\ .
\end{equation}

The above construction is actually a special case of a much more
general one. Let ${\cal N}_{\la\, x}{}^y$ be any NIM-rep, and let $J$ be
any simple current of the underlying fusion ring, say of order 
$m$. Then there will be a corresponding function \cite{SY}
$Q_J:P_+\rightarrow \frac{1}{m}{\Zop}$ such that 
\begin{equation}
S_{J\la,\mu}= \exp[2\pi{\rm i}\,Q_J(\mu)] \, 
S_{\la,\mu}\ .
\end{equation}
Then consider the $m$-fold cover 
where the boundary states are
labelled by $(x,j)$ (where $j\in \Zop_m$), and the NIM-rep is defined
by $\la.(x,j)=(\la.x,j+mQ_J(\la))$. This is really just the tensor
product  ${\cal N}\otimes \Zop_m$ of two NIM-reps. 
If $\mu\in{\cal E}$ are the exponents of the original NIM-rep  
${\cal N}$, then the exponents of the new NIM-rep will be the
multi-set $\cup_{i=0}^{m-1} J^i{\cal E}$. Now, we know that a NIM-rep
is decomposable (into a direct sum of smaller NIM-reps) if
and only if the multiplicity of the vacuum $0$ as an exponent is 
larger than one (see \cite{G}). Thus our new NIM-rep 
${\cal N} \otimes \Zop_m$ will be indecomposable, iff the old NIM-rep
${\cal N}$ was indecomposable and if no non-trivial power of the
simple current $J$ is itself an exponent of ${\cal N}$.

By these considerations, we know that our proposed NIM-rep ${\cal N} 
({\cal A}^*)\otimes \Zop_d$ has the correct exponents. Furthermore, it
is indecomposable since no non-trivial power of $J^{d'}$ is an
exponent of ${\cal N}({\cal A}^*)$. Thus it follows that the above
construction can be taken to define the NIM-rep 
${\cal N}({\cal D}^*)$.   

\subsection{Charges and charge groups}

Given that we have such a simple explicit formula for the NIM-rep, we
can analyse the corresponding charges in detail. These are 
integers $q(\widehat{a},i)$ satisfying
\begin{equation}\label{2.A}
{\rm dim}(\la)\,q(\widehat{a},i)= 
\sum_{\widehat{b}}{\cal N}({\cal A}^*)_{\la\, \widehat{a}}
{}^{\widehat{b}} \;\; q(\widehat{b},i+t(\lambda))\quad
({\rm mod}\ M)
\end{equation} 
for $M$ in (\ref{M_A}). The simplest solution to (\ref{2.A}) arises if
all $q(\widehat{0},i)$ are equal (for all $0\le i<d$). Then by an easy
induction argument from (3.4), (3.9) of \cite{Gaberdiel:2003kv}, we
get that {\it any}  $q(\widehat{a},i)$ is independent of $i$,  
and equals a solution $q_{{\cal A}^*}(\widehat{a})$ to the charge
equations for ${\cal A}^*$, and so by uniqueness (section 6.2 of 
\cite{Gaberdiel:2003kv}) $q(\widehat{a},i)$ equals the
common value of  $q(\widehat{0},i)$, times the dimension dim$_C(a)$
of the $C_m$-representation with highest weight $a=(a_1,\ldots,a_m)$
(see (3.1) of \cite{Gaberdiel:2003kv}). For $n$ odd, this uniqueness 
argument depends on a (safe) conjecture (see the discussion in section
1.2 of \cite{Gaberdiel:2003kv}). 

\noindent Thus it follows that ${\cal D}^*$ inherits a charge
assignment from ${\cal A}^*$, namely 
$q(\widehat{a},i)={\rm dim}_C(a)$. 
In particular, the charge group for ${\cal D}^*$ contains therefore
a summand $\Zop_{M}$. By analogy with the ${\cal D}$ charge
analysis of \cite{Gaberdiel:2004yn}, we should expect other 
solutions $q(\widehat{a},i)$ to (\ref{2.A}). However, any of these are
uniquely determined by the `initial values'  $q(\widehat{0},i)$.  
The reason is that if two solutions $q(\widehat{a},i)$,
$q'(\widehat{a},i)$ agree at $(\widehat{0},i)$, then by the argument
of the previous paragraph their difference must equal $0$ everywhere. 

In order to analyse the full charge group we therefore have to study
the solutions that are determined by all possible sets of `initial
values'. One constraint on these can be easily found as follows. 
All exponents $\mu$ of the  ${\cal A}^*$ NIM-rep obey
$\mu=\mu^*$. This implies that the matrices 
${\cal N}({\cal A}^*)_\la$ and ${\cal N}({\cal A}^*)_{\la^*}$ must be
equal (they have the same eigenvectors, namely the columns of
$\widehat{S}$, and the same eigenvalues, namely
$\tfrac{S_{\la^*\mu}}{S_{0\mu}}=
\tfrac{S_{\la\mu^*}}{S_{0\mu}}=\tfrac{S_{\la\mu}}{S_{0\mu}}$). Also, 
$t(\la^*)= -t(\la)$ (mod $n$) and 
dim$(\la^*)={\rm dim}(\la)$. Hence replacing $\lambda$ with $\lambda^*$
in (\ref{2.A}) yields
\begin{equation}
{\rm dim}(\la)\,q(\widehat{a},i)=\sum_{\widehat{b}}
{\cal N}({\cal A}^*)_{\la\widehat{a}}{}^{\widehat{b}}\;\;
q(\widehat{b},i-t(\la))
\quad({\rm mod}\ M)\ ,
\end{equation}
and comparing this with (\ref{2.A}) gives our basic constraint 
\begin{equation}\label{2.B}
{\rm dim}(\la)\,q(\widehat{0},i)= 
{\rm dim}(\la)\,q(\widehat{0},i+2t(\la))\quad
({\rm mod}\ M)\ ,
\end{equation}
valid for any $\la\in P_+^k({\rm su}(n))$ and $0\le i<d$. Of course,
(\ref{2.B}) holds for any $\widehat{a}$, but $\widehat{0}$ is all we need.

\subsubsection{The example of SU$(9)/\Zop_9$ at level $k=18$}

Since the argument in the general case is slightly involved it may be
instructive to see how one can use this equation to determine the
charge group in an explicit example. The example we want to study is 
SU$(9)/\Zop_9$ at level $k=18$. In this case $M=9$, and we want to
solve
\begin{equation}
{\rm dim}(\lambda)\, q(\hat{a},i)= \sum_{\hat{b}}  
{\cal N}_{\lambda\hat{a}}{}^{\hat{b}}\, 
q(\hat{b},i+t(\lambda))\quad ({\rm mod}\ 9) \ ,
\end{equation}
where ${\cal N}_{\lambda\hat{a}}{}^{\hat{b}}$ is the ${\cal A}^\ast$
NIM-rep. We now claim: 

\noindent {\bf Claim 1:} $q(\hat{a},i)={\rm dim}(a) Q_i$ is always  
a solution (mod 9), for any choice of $0\le Q_i<9$, $i=0,1,\ldots,8$,
provided only that $Q_i= Q_{i+3}= Q_{i+6}$ (mod 3), for 
$i=0,1,2$. (Note that this constraint follows from 
(\ref{2.B}) with $\lambda=\Lambda_3$ and $\lambda=\Lambda_6$.)
\smallskip

\noindent Let us assume Claim 1 for a moment. Then it is not
difficult to see that the charge group is 
\begin{equation}\label{su9ans}
{3\cdot\Zop_9 \oplus 6\cdot\Zop_3} 
\end{equation}
({\it i.e.}\ 3 copies of $\Zop_9$ and 6 of $\Zop_3$).
Indeed, Claim 1 implies that $Q_0,Q_1,Q_2$ are unconstrained, apart
from being between $0$ and $9$ (so they give us the $3\cdot\Zop_9$); then
the value of say $Q_3$ is determined (mod 3) by $Q_0$, so our  only 
freedom is to add $3$ or $6$ to that value (so each of those $Q_i$,
for  $3\le i<9$, contribute a copy of $\Zop_3$). 
\smallskip

\noindent So we want to prove Claim 1, {\it i.e.}\ that
\begin{equation}
{\rm dim}(\lambda)\, {\rm dim}_C(a) \,Q_i= 
Q_{i+t(\lambda)} \sum_{\hat{b}} 
{\cal N}_{\lambda\hat{a}}{}^{\hat{b}} \,
{\rm dim}_C(b)\quad ({\rm mod}\ 9)\ . 
\end{equation}
Of course, the ${\cal A}^*$ charge equation simplifies the right side,
and we  obtain
\begin{equation}\label{dimdim}
{\rm dim}(\lambda)\, {\rm dim}_C(a)\, Q_i= {\rm dim}(\la) \,{\rm dim}_C(a)\,
Q_{i+t(\lambda)}\quad ({\rm mod}\ 9)\ . 
\end{equation}
But dim$_C(a)$ are integers, and dim$_C(a)=1$ for
$a=0$. So (\ref{dimdim}) is equivalent to
\begin{equation}\label{2.s}
{\rm dim}(\lambda)\, Q_i= {\rm dim}(\la)\, Q_{i+t(\lambda)}\quad
({\rm mod}\ 9) \ . 
\end{equation} 
As we know, it is sufficient to consider only the generators $\la$ of
the fusion ring, {\it i.e.}\ the fundamental weights $\Lambda_j$ for  
$1\le j\le 8$. We also know dim$(\Lambda_j)$ is the binary coefficient
$({9\atop j})$. One can then easily check these fundamental weights
one by one to see that (\ref{2.s}) is indeed satisfied by the $Q_i$ in
Claim 1.

\subsubsection{The general argument}

Now we turn to the general argument. Recall from section~4.1 of 
\cite{Gaberdiel:2004yn} the following two facts:
\smallskip

\noindent{{\bf Proposition 1.}} \cite{Gaberdiel:2004yn} Consider any
su($n$) ($n$ can be even), and any prime power $p^\nu>1$ exactly
dividing $n$. Then
\begin{verse}
\noindent{(a)} gcd$_\la{\rm dim}(\la)={\rm gcd}_\ell
{\rm dim}(\Lambda_\ell) =p^\nu$, where we run over all weights \newline
$\la\in P_+({\rm su}(n))$
with $t(\la)$ coprime to $p$, and all $0<\ell<n$ coprime to $p$. 

\noindent{(b)} dim$(\Lambda_{p\ell})=
{\rm dim}(\bar{\Lambda}_\ell )$ (mod $p^\nu$), where $0<\ell<n/p$
is arbitrary, and $\Lambda_{p\ell}$ resp.\ $\bar{\Lambda}_\ell$
is a fundamental weight of su($n$) resp.\ su($n/p$).
\end{verse}
\medskip 

When we say $d$ exactly divides $n$, we mean $n/d$ is an integer
coprime to $d$. We write $d|n$ for a divisor and $d\| n$ for an exact
divisor. The dimension of a fundamental weight $\Lambda_j$ of su($n$)
is $\left({n\atop j}\right)$. We need to generalise part (a) of
this proposition. In the following, by  {\it e.g.} gcd$(d^\infty,n)$ 
we mean $\prod_{p|d}p^\nu$ where $p^\nu\|n$.
\medskip

\noindent{{\bf Proposition 2.}} Again let $n\ge 2$ be arbitrary. Then
\begin{verse}
\noindent{(a)} For
any divisors $d,e$ of $n$, with $e$ dividing $d^\infty$, define
$\bar{D}\equiv {\rm gcd}(d^\infty,n)$ and 
$\bar{D}_e\equiv{\rm gcd}_\la{\rm dim}(\la)$,  
where we run over all weights $\la\in P_+({\rm su}(n))$ with 
gcd$(t(\la),\bar{D})=e$. Then
$\bar{D}_e=\bar{D}/e$.

\noindent{(b)} Let $d$ be any divisor of $n$. Then gcd$_m{\rm dim}
(\Lambda_m)$, as $m$ ranges over all numbers coprime to $d$, is
gcd$(d^\infty,n)$.
\end{verse}\medskip

\noindent The proof of Proposition 2(a) is slightly involved; it is
given in appendix~A. The proof of Proposition 2(b) is a
straightforward simplification of that argument.

For any fixed $0\le i<j<d$, we defined $e=\,$gcd$(d,j-i)$. Then (at
least when $n$ is odd) eq.(\ref{2.B}) tells us 
\begin{equation}
{\rm dim}(\la)\,\Bigl(q(\widehat{0},i)-q(\widehat{0},j)\Bigr)= 0\quad 
({\rm mod}\ M) 
\end{equation}
for all $\la$ with gcd$(t(\la),d)=e$. Unfortunately, in 
Proposition 2(a) we have a slightly different gcd condition on the
$t(\la)$, and it can make a difference. For each prime $p$ dividing
$e$, let $p^\epsilon\|e$ as before, and put $\epsilon'=\epsilon$
unless $p^\epsilon\|d$, in which case put $\epsilon'= \nu$. 
(Recall that $p^\nu\|n$.)
Now replace $e$ with its multiple
$e'\equiv\prod_{p|e}p^{\epsilon'}$. When $d=\bar{D}$ then $e'=e$;
otherwise $e'$ may be larger than $e$.

Now we can use Proposition 2(a). For fixed $0\le i < j < d$ we set
$e=\,$gcd$(d,j-i)$ and define $e'$ as explained in the
previous paragraph. 
If $\lambda$ has the property that gcd$(t(\lambda),d)=e$ then
gcd$(t(\lambda),\bar{D})$ divides $e'$, and will equal $e'$ for some $\la$.
Thus Proposition 2(a) implies that for each such $i<j$ 
\begin{equation}\label{2.D}
q(\widehat{0},i)= q(\widehat{0},j)\quad
({\rm mod}\ M/D_{e'})\ ,\end{equation}
where $D_{e'}\equiv{\rm gcd}(\bar{D}/e',M)$.
\medskip

\noindent{\bf Result:} The general charge solution (\ref{2.A}), for
$n$ odd, for ${\cal D}^*$  is
\begin{equation}\label{nodres}
q(\widehat{a},i)={\rm dim}_C(a)\,
\left(Q+{M\over D}Q_i\right)\ ,
\end{equation}
valid with $M$ in (\ref{M_A}). Here $D={\rm gcd}(\bar{D},M)$,
and $0\le Q<M$ and $0\le Q_i<D$ are arbitrary except that
$Q_0=0$ and $Q_i= Q_j$ (mod $D/D_{e'}$), where $e'$ depends on $i<j$
as explained above. Hence the charge group for ${\cal D}^*$ is (for
odd $n$)   
\begin{equation}\label{noddres}
\boxed{
\Zop_M\oplus \bigoplus_{p|{\rm gcd}(d,M)}\; 
\bigoplus_{i=1}^\delta({p^i-p^{i-1}})\cdot\Zop_{p^{{\rm min} 
\{\mu,\nu-i+1\}}} }
\end{equation}
where $p^\nu\|n$, $p^\delta\|d$ and $p^\mu\|M$. The first sum runs
over all primes $p$ dividing both $d$ and $M$.\medskip 

For example, when $d$ is coprime to $M$, we get a charge group of
$\Zop_M$. When $d$ is a prime $p$ dividing $M$, then the charge group
is $\Zop_M\oplus (p-1)\cdot\Zop_D$. For SU$(9)/\Zop_9$ at $k=18$, we get
the charge group $3\cdot\Zop_9 \oplus 6\cdot\Zop_3$, as in the analysis of the
previous section.

The structure of this charge group for ${\cal D}^*$ ($n$ odd) is that
there is a `constant' solution ({\it i.e.}\ one independent of $i$),
defined mod $M$ (this accounts for the left-most summand $\Zop_M$); in
addition, there are 
solutions depending on $i$, and they are uniquely determined by the
`vacuum' values $q(\widehat{0},i)$ subject only to the
constraint (\ref{2.B}). $Q_i$ in (\ref{nodres}) gives the adjustment of
$q(\widehat{a},i)$ from $q(\widehat{a},0)$,
which is why we take $Q_0=0$. Eq.(\ref{noddres}) builds up the charge
group prime by prime; the $i$th summand there concerns the value of 
$Q_{j}$ for $p^{\delta-i}\le j<p^{\delta-i+1}$.
\medskip

The proof of `Result' is now easy (though as mentioned earlier it 
assumes Conjecture B of \cite{Gaberdiel:2003kv}). Eq.~(\ref{2.D})
gives an upper bound for the charge group, because 
$q(\widehat{0},i)$ uniquely 
determines the charge assignment. To see that the proposed charges
satisfy (\ref{2.A}), use the fact that dim$_C(a)$ solve the charge
equations for ${\cal A}^*$ to reduce (\ref{2.A}) to 
\begin{equation}\label{2.E}
{\rm dim}(\la)\,{\rm dim}_C(a){M\over D}Q_i={\rm dim}(\la)\,
{\rm dim}_C (a)\,{M\over D}Q_{i+t(\la)}\quad({\rm mod}\ M)\ .
\end{equation}
Let $e={\rm gcd}(d,t(\la))$, then $D_{e'}$ divides dim$(\la)$ by 
Proposition 2(a), and eq.(\ref{2.E}) is satisfied because of the
constraint on the $Q_i$. 

The Chinese Remainder Theorem permits us to treat distinct primes  
independently, {\it i.e.}\ we are only interested in the independent
values of each $Q_i$, modulo each sufficiently large prime power. So
fix a prime $p|d$, and write $p^\alpha\|D$. Then relation (\ref{2.D})
says that, modulo $p^\alpha$, the $Q_i$, for $0\le i<p^\delta$,
determine all remaining $Q_j$, $p^\delta \le j<d$, since $D_{e'}$ is
coprime to $p$ when $p^\delta$ divides $e$. So those $Q_j$ can be
ignored, as far as $p$ is concerned. Also, the $Q_i$, 
for $0\le i<p^{\delta-1}$, determine the $Q_j$, for 
$p^{\delta-1}\le j< p^\delta$, up to an ambiguity of
$\Zop_{D_p}=\Zop_{p^{{\rm min} \{\mu,\nu+1-\delta\}}}$. This
contributes 
$({p^\delta-p^{\delta-1}})\cdot\Zop_{p^{{\rm min}\{\mu,\nu+1-\delta\}}}$
to the charge group --- one $\Zop_{D_p}$ for each $j$. Continuing in
this way, we get the desired expression for the charge group. 

Since $\Zop_a\oplus\Zop_b\cong\Zop_{ab}$ whenever $a$ and $b$
are coprime, the ${\cal D}^*$ charge group can be rewritten in the
form (\ref{chgp}), where $M_1=M$, $t=d$, and $M_2,\ldots,M_d$ all 
divide gcd$(M,d^\infty,n)$.
In fact, most of these $M_i$ will usually equal 1: the number of
non-trivial $\Zop_{M_i}$ will be max$_{p|{\rm gcd}(d,M)}p^\delta$.

\section{The ${\cal D}^*$ case with $n$ even, non-pathological} 

In the previous section we computed the NIM-rep, charges and charge
group for the ${\cal D}^*$ NIM-rep when $n$ is odd. Even $n$ is 
more subtle. In particular, we learnt in \cite{Gaberdiel:2004yn} that 
(as far as ${\cal D}$ is concerned) there are two cases to
distinguish: $d'\equiv n/d$ even, which behaves almost as simply as
$n$ odd did, and is called {\it non-pathological}; and $d'$ odd, which
behaves much more peculiarly and is called {\it pathological}. In the
case of ${\cal D}$, the pathological behaviour was ultimately due to
dim$(J^{d'})= -1$ (mod $M$); for ${\cal D}^*$ this distinction remains
significant, though there is a different combinatorial cause.

In this section, we try to mimic the analysis of the previous section,
but for non-pathological even $n$. So $d$ can be any divisor of $n$,
provided $d'$ is even. We will find that we can be nearly as
successful here as we were for odd $n$.  

\subsection{The NIM-rep}

Write $n=2m$. The NIM-rep for $d=1$ ({\it i.e.}\ 
${\cal N}({\cal A}^*)$) was given in section~4.1.2 of
\cite{Gaberdiel:2002qa}.  The exponents of ${\cal A}^*$ consist of all
$\mu\in P_+^{k}({\rm su}(n))$ with
$C\mu=\mu$, all with multiplicity 1. The boundary states $\widehat{a}$
are parametrised by the level $k$ integrable highest weights
of the twisted affine algebra $A_{2m-1}^{(2)}$, {\it i.e.}\ they are all
$(m+1)$-tuples $(a_0;a_1,\ldots,a_m)$ of non-negative integers where
$k=a_0+a_1+2a_2+ \cdots+2a_m$. The $\psi$-matrix, diagonalising the
NIM-rep, is the modular $\widehat{S}$-matrix of $A_{2m-1}^{(2)}$ (see
eq.~(4.5) in \cite{Gaberdiel:2002qa}). The NIM-rep coefficients are
$C$-twisted fusion coefficients of $A_{n-1}^{(1)}$ level $k$, 
and can be expressed in terms of ordinary fusions of 
$B_{m}^{(1)}$ level $k+1$  \cite{Gaberdiel:2003kv}. Thanks to the
order-2 symmetry 
of the  Dynkin diagram of the `orbit Lie algebra' $D_{m+1}^{(2)}$, the
NIM-rep has a grading: 
\begin{equation}\label{3.F}
{\cal N}({\cal A}^*)_{\la\widehat{a}}{}^{\widehat{b}}\ne 0\ 
\Longrightarrow
\ t(\la)+Q(\widehat{a})= Q(\widehat{b})\quad ({\rm mod}\ 2) \ ,
\end{equation}
where $t(\la)=\sum_{i=1}^{n-1}i\la_i$ as usual and 
$Q(\widehat{a})= \sum_{i=1}^{m}ia_i$. Because of the order-2 symmetry
of the Dynkin diagram of the twisted algebra $A_{2m-1}^{(2)}$, the
NIM-rep has a symmetry: 
\begin{equation}\label{3.G}
{\cal N}({\cal A}^*)_{\la\widehat{a}}{}^{\widehat{b}}=
{\cal N}({\cal A}^*)_{\la,K\widehat{a}}{}^{K\widehat{b}}\ ,
\end{equation}
where $K(a_0;a_1,a_2,\ldots,a_m)=(a_1;a_0,a_2,\ldots,a_m)$; see
section~5.3 of \cite{Gaberdiel:2002qa} for more details. These are
special features 
of $n$ being even, and will play a crucial role in the following. 

We found in \cite{Gaberdiel:2003kv} that the charge group for ${\cal
  A}^*$ is 
$\Zop_{M}$, just as it is for ${\cal A}$, and the charge
$q(\widehat{a})$ can be taken to be dim$_C(a)$. These facts, 
which we need below, could only be proven once we assumed Conjectures B
and B$^{{\rm spin}}$ (see section~1.2 of  \cite{Gaberdiel:2003kv} for
details). These conjectures seem quite safe. 

If $n$ and $d'$ are even the exponents of ${\cal D}^*$ form the 
multi-set $\bigcup_{j=1}^{d}\bigcup_{\mu=C\mu}J^{jd'/2}\mu$. That is,
$\mu\in P_+^{k}({\rm su}(n))$ will have multiplicity $0$ unless
$C(J^{jd'/2}\mu)=J^{jd'/2} \mu$ for some $j$, in which case its
multiplicity will equal its order as a $J^{d'}$-fixed point. The $1/2$
in the exponent was necessary in order to move each exponent into a
$C$-invariant weight. This is a key difference between pathological
and non-pathological cases: in the pathological case 
({\it i.e.}\ $d'$ odd),
not all exponents are related to $C$-invariant weights, and so
there is no chance of a direct relation between ${\cal N}({\cal D}^*)$
and ${\cal N}({\cal A}^*)$. The reason we take $1\le j\le d$ in the
union rather than $1\le j\le 2d$ is because otherwise we would 
double-count: if $C\mu=\mu$, then also $C(J^{n/2}\mu)=J^{n/2}\mu$. 

The previous construction ${\cal N}({\cal A}^*)\otimes \Zop_d$ fails
here (at least for $d$ even), because $J^{n/2}0$ is an exponent for 
${\cal N}({\cal A}^*)$, so $0$ is an exponent of ${\cal N}({\cal A}^*)
\otimes \Zop_d$ with multiplicity 2 (when $d$ is even). This means that 
${\cal N}({\cal A}^*)\otimes\Zop_d$ is decomposable into a direct sum
of $2$ (isomorphic) NIM-reps (when $d$ is even). More
precisely, thanks to the grading (\ref{3.F}), the action 
\begin{equation}\label{3.3}
\la.(\widehat{a},i)=(\la.\widehat{a},t(\la)+i)
\end{equation}
obeys the property 
\begin{equation}
Q(\la.\widehat{a})+(t(\la)+i) = Q(\widehat{a})+i \quad
{\rm (mod}\ 2) \ . 
\end{equation}
Thus the two (indecomposable) summands in ${\cal N}({\cal A}^*)
\otimes\Zop_d$ (for $d$ even) consist of the pairs $(\widehat{a},i)$
with $Q(\widehat{a})$ and $i$ congruent, or not congruent, mod $2$. 

Instead, the correct ${\cal D}^*$ NIM-rep is an irreducible summand of
${\cal N}({\cal A}^*)\otimes \Zop_{2d}$, for definiteness, say, the
one obeying
\begin{equation}\label{nevrest}
Q(\widehat{a}) = i\quad {\rm (mod}\  2) \ . 
\end{equation}
This is valid for $d$ even or odd, provided only that $d'$ is   
even. The proof is simple: the exponents of 
${\cal N}({\cal A}^*)\otimes \Zop_{2d}$ are $J^{jd'/2}\mu$ for 
all $C\mu=\mu$ and $0\le j<2d$, and the exponents of the irreducible
summand are half of those ({\it i.e.}\ divide each multiplicity by
$2$). Note again the importance of $d'$ being even: 
${\cal N}({\cal A}^*)\otimes  \Zop_{2d}$ will define a NIM-rep only if
2$d$ divides $n$. 

When $d$ is odd, there is a more direct construction of this NIM-rep:
it is simply ${\cal A}^*\otimes\Zop_d$. The isomorphism sends 
$(\widehat{a},i)\in {\cal A}^*\otimes\Zop_d$ to
$(\widehat{a},dQ(\widehat{a})+(1-d)i)\in {\cal A}^*\otimes\Zop_{2d}$. 
Then 
\begin{eqnarray}
\la.\Bigl(\widehat{a},dQ(\widehat{a})+(1-d)i\Bigr) 
& = & 
\Bigl(\la.\widehat{a},dQ(\widehat{a})+(1-d)i+t(\la)\Bigr) \nonumber 
\\
& = & \Bigl(\la.\widehat{a},dQ(\la.\widehat{a})+(1-d)(i+t(\lambda))
\Bigr)\ , 
\end{eqnarray}
which indeed corresponds to $(\la.\widehat{a},i+t(\la))$, and so the
two $\la$-actions agree. On the other hand,  when $d$ is 
even, $1-d$ is invertible mod $2d$ so this merely defines an automorphism
of ${\cal N}({\cal A}^*)\otimes\Zop_{2d}$, rather than a projection onto 
${\cal N}({\cal A}^*) \otimes\Zop_d$. 

The boundary states of ${\cal D}^*$ can thus be identified with pairs 
$(\widehat{a},i)$ where $Q(\widehat{a})= i$ (mod 2).
The $\psi$ matrix for the NIM-rep ${\cal N}({\cal A}^*)\otimes\Zop_{2d}$
is given as in the
previous section, with  $d$ replaced everywhere with $2d$. The
$\psi$-matrix for ${\cal D}^*$ is then 
\begin{equation}
\psi({\cal D}^*)_{(\widehat{a},i),(\mu,j)}={1\over \sqrt{d}}
\exp[\pi{\rm i}\,ij/d]\widehat{S}_{\hat{a},\mu} \ , 
\end{equation}
where $\mu=C\mu$, $0\le i<2d$, $i= Q(\widehat{a})$ (mod 2), and 
$1\le j\le d$. The restriction of the rows ({\it i.e.}\ of $i$) is
clear; the  restriction of the columns is because the
$(K\widehat{a},j)$ column now equals the $(\widehat{a},j+d)$
column. As before, this amounts to specifying the exponents of
${\cal D}^*$ by pairs $(\mu,j)$, where $\mu=C\mu$ and $1\le j\le d$.  

\subsection{Charges and charge groups}

Having given a fairly explicit description of ${\cal N}({\cal D}^*)$
for the non-pathological ($d'$ even) case for $n$ even, we can now
attempt to determine the corresponding charge group. As we have seen
above, the structure of the NIM-rep depends on whether $d$ is even or
not.  

If $d$ is odd, then ${\cal N}({\cal D}^*)$ has in fact the same
structure as for $n$ odd, and the charges and charge group is given as
in the `Result' of the previous section. Indeed, the identical
arguments apply. 

If $d$ is even, on the other hand, the structure of the NIM-rep is
more complicated, and so is the analysis of its charges.
As in section~2.2, the charges are uniquely determined by the values
$q(\widehat{0},i)$. Also, taking $q(\widehat{a},i)={\rm dim}_C(a)$
defines an order $M$ solution. Our proof of 
the Result in section~2 followed because constraint (\ref{2.B}) there
gives an upper bound for the charge group, while the charge ansatz 
(\ref{nodres}) gives a lower bound, and the two agree. Constraint
(\ref{2.B}) now implies eq.~(\ref{2.D}) as before; although the $2$ in
(\ref{2.B}) is no longer invertible mod $d$, it is now superfluous
because of the parity condition on the $i$. However the ansatz
(\ref{nodres}) now in general works only mod $M/2$.  

\noindent That ansatz does not respect (\ref{nevrest}); a more natural 
one is to introduce parameters $Q_{i,s}$ for $0\le i<2d$ and $s=0,1$
where
\begin{equation}\label{3.H}
q(\widehat{a},i)={\rm dim}_C(a)\,{M\over D} 
Q_{i,Q(\widehat{a})}\ .
\end{equation}
Now we can demand $Q_{i,s}=0$ unless $i=s$ (mod 2). Unfortunately, this
ansatz fairs no better: it solves the charge equation provided we take
$Q_{i,0}= Q_{j,0}$ (mod $D/D_{e'}$), where we restrict to even $i,j$,
and define $e'$ using $2d$ rather than $d$. The result is that the 
${\cal D}^*$ charge group contains
\begin{equation}\label{nevlow}
\Zop_M\oplus \bigoplus_{i=1}^\delta({2^i-2^{i-1}})\cdot
\Zop_{2^{{\rm min}
\{\mu,\nu-i\}}}\oplus\bigoplus_{{p|{\rm gcd}(d,M)\atop p\ne 2}}
\bigoplus_{i=1}^\delta({p^i-p^{i-1}})
\Zop_{p^{{\rm min}\{\mu,\nu-i+1\}}} \ .
\end{equation}
We also know from (\ref{2.D}) that it is contained in the charge group
(\ref{noddres}). 

The ${\cal D}^*$-charge group is a {\it proper} subgroup of
(\ref{noddres}) iff (\ref{2.B}) can be supplemented by new constraints
on $q(\widehat{0},i)$; it {\it properly} contains (\ref{nevlow}) iff
a more general ansatz than (\ref{3.H}) can be found and is
effective. When there are at least as many 2's dividing $n$ as $dM$,
the two bounds agree. 

We conjecture that the ${\cal D}^*$-charge group is {\it always} given
by (\ref{noddres}). While we do not have a general argument for this
assertion, we have checked it explicitly for the simplest non-trivial
case SU$(4)/\Zop_2$ (see below). Also, if the ${\cal D}^*$-charge
group is to agree with the ${\cal D}$-charge group (as it appears to
do), then the analysis of the next section implies that the 
${\cal D}^*$-charge group must be (\ref{noddres}). We regard this as 
convincing evidence that the ${\cal D}^*$-charge group is indeed
given by (\ref{noddres}).

For the example of SU$(4)/\Zop_2$ the two bounds (\ref{noddres}), 
(\ref{nevlow}) agree, and therefore equal the ${\cal D}^*$-charge
group, unless $k$ is an odd multiple of 4, so it suffices to restrict
attention here to the latter. The boundary labels are $([a_1,a_2],j)$
where $0\le j<4$, $a_i\ge 0$, and $a_1+2a_2\le k$. The NIM-rep is
built out of the ${\cal A}^*$ one in the way described above; the
${\cal A}^*$ NIM-rep is generated by 
\begin{eqnarray}
\Lambda_1.[a_1,a_2]&=&\Lambda_3.[a_1,a_2]=[a_1+1,a_2]+[a_1-1,a_2]+
[a_1-1,a_2+1]+[a_1+1,a_2-1] \nonumber \\
\Lambda_2.[a_1,a_2]&=&2[a_1,a_2]+[a_1,a_2-1]+[a_1-2,a_2+1]+[a_1+2,a_2-1]
+[a_1,a_2+1]\ , \nonumber
\end{eqnarray}
where we drop any term $[a_1',a_2']$ with a component $a'_1,a'_2$, or 
$a'_0\equiv k-a_1'-2a'_2$ equal to $-1$. In addition, when $a_1=0$ use
$[-2,a_2+1]=-[0,a_2]$ and when $k=a_1+2a_2$ use
$[a_1,a_2+1]=-[a_1,a_2]$. 

We have checked by an explicit computation that for all such $k$ 
the charge group is $\Zop_M\oplus \Zop_4\oplus\Zop_4$,  
in agreement with (\ref{noddres}). The solution of order $M$ is simply
$q(\widehat{a},j)={\rm dim}_C(\widehat{a})$; the order $4$ assignments
are distinguished by their values of $q(\widehat{0},0)$ and
$q(\widehat{0},2)$, and have the special property that the charges
$q([a_1,a_2],j)$ are periodic in both $a_1,a_2$ with period
$8$. (Given this periodicity, one can check the existence of this
solution fairly straightforwardly for all such $k$.)

\section{The ${\cal D}$ charge group in the non-pathological case}  

We now want to compare the results of the previous two sections with
the  charge analysis for the ${\cal D}$ theory. Partial results for
that were already found in \cite{Gaberdiel:2004yn}, but now we are
able to go several steps further, at least in the non-pathological
case.

\subsection{The NIM-rep}

The modular invariant here is given by (\ref{modular}), and requires
that 
$n(n+1)k/d$ be even. Write $f={\rm gcd}(d,k)$. Explicitly listing
the exponents requires distinguishing two cases:
\smallskip  

\noindent{{\bf Case A:}} Either $n$ or $k$ is odd, or either $n/d$ or
$k/f$ is even. Then $\mu\in P_+^k({\rm su}(n))$ is an exponent if and
only if $d$ divides $t(\mu)$. Such a $\mu$ has multiplicity
$o(\mu)$. 
\smallskip

\noindent{{\bf Case B:}} Both $n$ and $k$ are even, and both 
$n/d$ and $k/f$ are odd. Then the exponents come in two versions
(see \cite{Gaberdiel:2004yn} for details). Case B is necessarily 
pathological, and we shall not consider it in this paper.
\smallskip 

The boundary states $a\in{\cal B}$ here correspond to pairs 
$([\nu],i)$, where $[\nu]=\{J^{j d'}\nu\}$ is the
$J^{d'}$-orbit of any weight $\nu\in P_{+}^{k}({\rm su}(n))$, and where
$1\le i\le o(\nu)$. To simplify notation, we will write $[\nu,i]$ 
for $([\nu],i)$, and when $\nu$ has order $o(\nu)=1$, then we shall
usually write $[\nu]$ instead of $([\nu],1)=[\nu,1]$. 

We will restrict attention in this paper to Case A: it contains all
non-pathological cases, as well as some pathological ones. The
$\psi$-matrix is given by  (\ref{A.4}), and the NIM-rep 
${\cal N}({\cal D})$ is thus obtained from (\ref{nimpsi}). More
explicitly, it can be described as follows. 
When $\nu$ is not a fixed point, (1.17) of \cite{Gaberdiel:2004yn}
reads  
\begin{equation} \label{nimfus}
\N({\cal D})_{\la\, [\nu]}{}^{[\nu',i]}=
\sum_{j=1}^{d/ o(\nu')}N_{\la\, \nu}{}^{J^{d'j}\nu'}
\end{equation}
for any weight $\la\in P_+^k({\rm su}(n))$ and boundary label $[\nu',i]$.
More generally, for any weight $\la\in P_{+}^{k}({\rm su}(n))$ and
boundary labels  $[\nu,i],[\nu',i']$, (1.18) of
\cite{Gaberdiel:2004yn} reads 
\begin{equation}\label{nimsum}
\sum_{i=1}^{o(\nu)}\N({\cal D})_{\la\, [\nu,i]}{}^{[\nu',i']}=
\sum_{j=1}^{d/ o(\nu')}N_{\la\, \nu}{}^{J^{d'j}\nu'}\ .
\end{equation}

Computing the remaining NIM-rep entries is much more subtle, but
in appendix~B we obtain these for the fundamental weights and simple
currents. (Recall that the NIM-rep matrices for
the fundamental weights determine all NIM-rep matrices
uniquely.) The result is an unexpectedly simple generalisation of
(\ref{nimfus}):
\begin{eqnarray}
\N({\cal D})_{\Lambda_m,[\varphi,i]}{}^{[\varphi',i']}& =&\,
\sum_{j=1}^{d/g''}
\delta_{ii'}^{(\Delta)}N^{(\Delta)}_{J^{d'j}\Lambda_{m/\Delta},
\varphi^\Delta}{}^{\varphi'{}^\Delta} \label{4.8a}\\
\N({\cal D})_{J0,[\varphi,i]}{}^{[\varphi',i']}&=&\,
\delta_{[\varphi'],[J\varphi]} \delta_{i,i'} \label{4.8b} \ , 
\end{eqnarray} 
where again we interpret the right-side of 
(\ref{4.8a})
as vanishing if $\Delta$ does not divide $m$. Here, 
$g''={\rm lcm}(o(\varphi),o(\varphi'))$ and 
$\Delta={\rm gcd}(o(\varphi),o(\varphi'))$. $N^{(\Delta)}$ stands for
the  fusion coefficients for 
$\widehat{\rm su}(n/\Delta)_{k/\Delta}$. $\varphi^\Delta$ means to  
truncate the $n$-tuple $\varphi$ after $n/\Delta$ components, and
to regard it as a weight of su$(n/\Delta)$. We verify these formulae
in appendix~B, using the fixed point factorisation formulae of
\cite{GW}. This new formula (\ref{4.8a}) is why we can now say much
more about the ${\cal D}$-charges than we could in
\cite{Gaberdiel:2004yn}.

\subsection{Charges and charge groups}

With our improved understanding of the NIM-rep for ${\cal D}$,
we can now do much better at determining its charge group than in the
original analysis of \cite{Gaberdiel:2004yn}. For concreteness we will
first discuss the case of SU$(9)/\Zop_9$ at level $k=18$.

\subsubsection{The example of  SU$(9)/\Zop_9$ at level $k=18$}

The new key ingredient are the `fixed point factorisation' formulae,
expressing the ${\cal D}$-NIM-rep at fixed points in terms of
$\widehat{{\rm su}}(3)_6$ fusions. The interesting NIM-rep
entries are (\ref{4.8a}):
\begin{itemize}
\item[(i)] when $3|m$ and the order of fixed points $\phi,\psi$ are
both $3$, then  
\begin{equation} \label{su9fpfa}
\N({\cal D})_{\Lambda_m,[\phi,i]}{}^{[\psi,j]}=\sum_{h=1}^3
\delta_{ij} \, \bar{N}_{\bar{J}^h\bar{\Lambda}_{m/3},\bar{\phi}}
{}^{\bar{\psi}} \ .
\end{equation}
\item[(ii)] when $3|m$ and the order of fixed points $\phi,\psi$ are 
$3$ and $9$, then  
\begin{equation} \label{su9fpfb}
\N({\cal D})_{\Lambda_m,[\phi,i]}{}^{[\psi,j]}=\delta_{ij}^{(3)} \, 
\bar{N}_{\bar{\Lambda}_{m/3},\bar{\phi}}{}^{\bar{\psi}} \ . 
\end{equation}
\end{itemize}
We let bars denote $\widehat{{\rm su}}(3)_6$ quantities.
We know already from \cite{Gaberdiel:2004yn} that the `unresolved'
charge group for ${\cal D}$ is $\Zop_9$. So we only need to determine
the `resolved' charge group.\footnote{In \cite{Gaberdiel:2004yn}
we called the `unresolved' and `resolved' charge groups `untwisted'
and `twisted', respectively. In the current context this is bound to
lead to confusions! We shall therefore use the terms 
`unresolved' and `resolved' in this paper.}
That is, we can impose the condition that
the charges $q[\lambda]$ of any non-fixed point $\lambda$ be 
$0$. The remaining charges can be parametrised by su(3) weights ---
more precisely by pairs $[\bar{\phi},i]\equiv ([\bar{\phi}],i)$, where
$[\bar{\phi}]$ is a $\bar{J}$-orbit in $\widehat{\rm su}(3)_6$, and 
$1\le i\le 3$ if $\bar{\phi}$ is not the fixed point $(22)$, while
$1\le i\le 9$ if $\bar{\phi}=(22)$. We need to understand the complete
list of constraints on these `resolved' charges $q[\bar{\phi},i]$. 

One constraint is easy.
Let $\phi$ be the $\widehat{\rm su}(9)_{18}$ fixed point
$(\bar{\phi},\bar{\phi},\bar{\phi})$ corresponding to the
$\widehat{\rm su}(3)_6$ weight $\bar{\phi}$, and write $\bar{0}$ for
$(0,0)$. Then we know from \cite{Gaberdiel:2004yn} (or directly from
the ${\cal D}(\widehat{\rm su}(9)_{18})$ charge equation  with
$\lambda=\phi$ and $a=[\bar{0},j]$) that 
\begin{equation}\label{4.s}
\sum_{j} q[\bar{\phi},j]=0\quad({\rm mod}\ 9)\ ,
\end{equation}
where we sum over $1\le j\le 3$ for $\bar{\phi}\ne (2,2)$, and over
$1\le i\le 9$ for $\bar{\phi}=(2,2)$.

We know on general grounds that it is sufficient to consider the
charge equations only for the fundamental weights, {\it i.e.}\ the
equations 
\begin{equation}\label{HS}
{\rm dim} (\Lambda_m) \, q[\lambda,i] = 
\sum_{[\mu,j]} \N_{\Lambda_m,[\lambda,i]}{}^{[\mu,j]} \, 
q[\mu,j] \quad({\rm mod}\ 9)\ . 
\end{equation}
 Because we are in the realm of the
`resolved' charges, we only have to worry about the charges of fixed 
points on the right-hand-side, and the relevant NIM-rep coefficients
are given by the fixed point factorisation formulae (\ref{su9fpfa})
and (\ref{su9fpfb}).  

If $\lambda$ is not an $\widehat{\rm su}(9)_{18}$ fixed point, then by
assumption $q[\lambda,i]=0$, so the right-hand-side sum must
also be $0$ (mod 9). But the relevant NIM-rep coefficients in this
case are given by  (\ref{nimfus}), and so in
particular the right-hand-side of that charge equation will involve
sums as in (\ref{4.s}). Thus the charge equation (\ref{HS}) when
$\lambda$ is not an $\widehat{\rm su}(9)_{18}$ fixed point are
automatically satisfied as long as (\ref{4.s}) is satisfied; they
therefore do not supply any new constraints. 

So all we have to consider are dim$(\Lambda_m) q[\bar{\phi},i]$.
If $3$ does not divide $m$, then 9 will divide dim$(\Lambda_m)$ (by
Proposition 1 --- see section~2.2.2), so the left-hand-side of
(\ref{HS}) will be $0$. But so  
will the right-hand-side since $3$ will not divide the  `triality' of
$\Lambda_m$ (namely $m$) plus the triality of $\phi$ (namely $3
t(\bar{\phi})+9k$), so no fixed points can appear on the 
right-hand-side. Thus the charge equation is trivially satisfied. 

So all we really have to consider are dim$(\Lambda_m) q[\bar{\phi},i]$
when $m=3$ or $m=6$. In this case the left-hand-side of (\ref{HS})
becomes $3 q[\bar{\phi},i]$ (mod 9). Using the fixed point
factorisation formulae (\ref{su9fpfa}) and (\ref{su9fpfb}), the
right-hand-side becomes  
\begin{equation}
\sum_{[\bar{\psi}]\ne (22)}\sum_{h=1}^3
\bar{N}_{\bar{J}^h\bar{\Lambda}_{m/3},\bar{\phi}}{}^{\bar{\psi}} 
q[\bar{\psi},i]+\bar{N}_{\bar{\Lambda}_{m/3},\bar{\phi}}{}^{(22)}
\sum_{h=1}^3 q[(22),i+3h] \quad ({\rm mod}\ 9) 
\end{equation}
provided that $\bar{\phi}\ne (22)$, and 
\begin{equation}
\sum_{[\bar{\psi}]\ne(22)}
\bar{N}_{\bar{J}^h\bar{\Lambda}_{m/3},\bar{\phi}}
{}^{\bar{\psi}} q[\bar{\psi},i]\quad ({\rm mod}\ 9)
\end{equation}
if $\bar{\phi}=(22)$. Here, we have assumed for notational convenience
that $q[\bar{\psi},i]$ is periodic in $i$ with period $3$ (for
$\bar{\psi}\ne (22)$) or $9$ (for $\bar{\psi}=(22)$).  

The point of all this is, that we are to recognise these NIM-rep 
coefficients (for \linebreak
${\cal D}(\widehat{\rm su}(9)_{18})$ involving $2$
fixed points) as precisely the NIM-rep coefficients for  
${\cal D}(\widehat{\rm su}(3)_6)$. That is, these last charge
equations are indistinguishable from those for 
${\cal D}(\widehat{\rm su}(3)_6)$!  

Collecting all this together, we obtain the statement that the
`resolved' charge equations for the ${\cal D}$ charge group of 
SU(9)$/\Zop_9$, $k=18$, consist of 
(\ref{4.s}), together with $3$ decoupled copies of the whole
collection of charge equations for the ${\cal D}$ charge group of 
SU(3)$/\Zop_3$, $k=6$. The $3$ decoupled equations apply to the $3$
values of $i$ (mod 3); the ${\cal D}$ charge equations for
SU(3)$/\Zop_3$ were analysed in detail in section~2 of 
\cite{Gaberdiel:2004yn}.  

Ignoring (\ref{4.s}) temporarily, this means we would get a 
net resolved charge group of $3$ copies of the ${\cal D}$ charge group
of $\widehat{\rm su}(3)_6$, {\it i.e.} 
$$
3\cdot(\Zop_9\oplus 2\cdot\Zop_3)= 3\cdot\Zop_9\oplus 6\cdot\Zop_3\ .
$$
The 3 $\Zop_9$'s correspond to the values of $q[\bar{0},i]$ 
($1\le i\le 3$); the 6 $\Zop_3$'s correspond to the values of 
$q[(22),j]$ ($1\le j\le 6$). More precisely, the values of
$q[\bar{0},i]$ can take any value from $0$ to $8$, whereas
$q[(22),j]$ are constrained by
(2.14) and (2.16) of \cite{Gaberdiel:2004yn}, which say that 
\begin{equation}\label{first}
q[(22),j]+q[(22),j+3]+q[(22),j+6]=0
\quad ({\rm mod}\ 9)
\end{equation}
and 
\begin{equation}\label{second}
3 q[(22),j] = 6 \quad ({\rm mod}\ 9)  \ ,
\end{equation}
respectively. The first equation (\ref{first}) fixes 
$q[(22),7]$, $q[(22),8]$, $q[(22),9]$ in terms of
$q[(22),j]$ with
$1\le j\le 6$. The second (\ref{second}) implies that 
$q[(22),j]=2$ (mod $3$), which fixes  the remaining 
$q[(22),j]$ up to a $\Zop_3$ ambiguity.   

This leaves us with imposing (\ref{4.s}). For $\bar{\phi}=(22)$ 
we get simply the constraint (\ref{first}) again. 
So this means that we keep the 6 $\Zop_3$'s. But (\ref{4.s}) for 
$\bar{0}$ is non-trivial, and fixes 
$q[\bar{0},3]=-q[\bar{0},1]-q[\bar{0},2]$. This drops
the $3\cdot\Zop_9$ to $2\cdot\Zop_9$. Then (\ref{4.s}) will be
automatically satisfied for the remaining $\bar{\phi}$, since
$q[\bar{\phi},i]$ will equal dim$(\bar{\phi})q[\bar{0},i]$.  

Thus the resolved ${\cal D}$ charge group for SU(9)$/\Zop_9$ level
$k=18$ is in fact $2\cdot\Zop_9\oplus 6\cdot\Zop_3$. Together with the
unresolved charge group $\Zop_9$, the total ${\cal D}$ charge group is
therefore  
\begin{equation}
3\cdot\Zop_9 \oplus 6\cdot\Zop_3 \ .
\end{equation}
This agrees precisely with the result for the ${\cal D}^*$ charge
group, eq.~(\ref{su9ans}). 

More generally, the identical argument
applies for SU(9)$/\Zop_9$ whenever $9$ divides the level $k$, in
which case we obtain the resolved charge group $2\cdot\Zop_M\oplus
6\cdot\Zop_3$. Together with the unresolved charge group $\Zop_M$,    
this means that the ${\cal D}$ and ${\cal D}^*$ charge groups for
SU(9)$/\Zop_9$  when 9 divides $k$, are both 
\begin{equation}
3\cdot\Zop_M\oplus 6\cdot\Zop_3 \ . 
\end{equation}

\subsubsection{The general argument}

For general $n$ (Case A), the same things happen: in particular, 
the resolved charge equations for SU$(n)/\Zop_d$ can be expressed in 
terms of the full charge equations of SU$(n/\Delta)/\Zop_{d/\Delta}$
for some $\Delta$, using fixed point factorisation. This 
means that the resolved charge group for SU$(n)/\Zop_d$ can be
expressed in terms of (un)resolved charge groups for 
SU$(n/\Delta)/\Zop_{d/\Delta}$, although a non-trivial amount of 
book-keeping has to be done to arrive at the final answer.

The full ${\cal D}$-charge group appears to be given by
(\ref{noddres}), but we do not have a general proof of this yet. The 
structure of this charge group says that there is a charge assignment 
given by $q([0])=1$ defined mod $M$ which accounts for the left-most 
summand $\Zop_M$; it corresponds to the `unresolved' charges (that
were called `untwisted' in \cite{Gaberdiel:2004yn}).
It obeys $q([\lambda])={\rm dim}(\la)$ on
non-fixed points, but is not uniquely defined on the fixed
points. That ambiguity is completely captured by what we call
the `resolved charges' (that were called `twisted charges' 
in \cite{Gaberdiel:2004yn} --- charge assignments 
with $q([0])=0$). We will find that resolved charges are more tractable
than unresolved ones. The charge assignments are uniquely
determined by the values of $q([0])$ and 
$q([0_{p^i}, j])$,
where $0_{p^i}=(k/p^i;0,\ldots,0,k/p^i,0,\ldots)$ with a
$k/p^i$ placed in every $n/p^i$th entry. 
Here $1\le i\le\delta$ with $p^\delta \| d$, and 
$1\le j\le p^i$. The fixed point $0_{p^i}$ 
generates all order $p^i$ fixed points, and corresponds to the vacuum 
in $\widehat{{\rm su}}(n/p^i)_{k/p^i}$. All possible charge
assignments can be explicitly built up inductively from those of the
prime-power orbifolds SU$(n/p^i)/\Zop_{p^{\delta-i}}$. In particular,
the summands in (\ref{noddres}) associated with $i$ correspond to 
the freedom in choosing the values of
$q([0_{p^i}, j])$.

We can prove most of these statements, as we shall see shortly. The
main uncertainty is the existence of an unresolved 
${\cal D}$-charge assignment for each SU($n/p^i)/\Zop_{p^{\delta-i}}$
level ${k/p^i}$. Our arguments hold for Case A (which includes all
non-pathological cases as well as some pathological ones). 
As (\ref{noddres}) indicates, we will want to look at each prime $p$
separately --- see appendix~C for the detailed arguments.
Consider first the resolved charges of the SU($n)/\Zop_d$ 
${\cal D}$-theory. As 
before we define for each prime $p$, the parameters $\nu$, $\mu$ and 
$\delta$ by  $p^\nu\|n$, $p^\mu\|M$ and $p^\delta\|d$. 
\smallskip 

The following congruences uniquely specify integers $q[\phi,i]$
mod $M$. For each prime $p$ dividing both $M$ and $d$,
let $q_p[\phi,i]$ be a resolved charge assignment (mod $M$) for 
SU$(n)/\Zop_{p^\delta}$ level $k$. For a given $J^{d'}$-fixed point
$\phi$ with order $o(\phi)$, let $p^\ell\|o(\phi)$. We require that
our integers $q[\phi,i]$ satisfy 
$q[\phi,i]=(o(\phi)/p^\ell)^{-1}q_p[\phi,i]$ 
(mod $p^\mu$) ($o(\phi)/p^\ell$ is coprime to $p$ so is invertible mod
$p^\mu$). For each prime $p$ dividing $M$ but not $d$, we also require
$q[\phi,i]=0$ (mod $p^\mu$). This fixes $q[\phi,i]$ uniquely, and it
is not difficult to see that the integers so obtained define a
resolved charge assignment for SU$(n)/\Zop_{d}$ level
$k$. Moreover, all resolved charge assignments for SU$(n)/\Zop_{d}$
can be described in this way. 

This tells us that for resolved charges, it suffices to consider 
orbifolds by prime powers. Note that the orbits $[\phi]$ in the
charges $q_p[\phi,i]$ defined for SU$(n)/ \Zop_{p^\delta}$ are with 
respect to $J^{n/p^\delta}$, while the orbits for the actual charges 
$q[\phi,i]$ in SU$(n)/\Zop_d$ are with respect to $J^{n/d}$. The
former are smaller, by a factor of $o(\phi)/p^\ell$, which is
precisely the origin of the above factor. Eq.(\ref{4.8b}) 
guarantees that $q_p[J^{d'}\phi,i]=q_p[\phi,i]$, so our formula is
well-defined.

It thus remains to construct resolved charge assignments for
SU$(n)/\Zop_{p^\delta}$ level $k$. Choose any $p$ ${\cal D}$-charge
assignments $q_p^{(h)}[\lambda_p,i]$ for 
SU$(n/p)/\Zop_{p^{\delta-1}}$ level $k/p$, where $1\le h\le p$,
$\lambda_p\in P_+^{k/p}({\rm su}(n/p))$, and $i$ runs from 1 to the
$J^{n/p^{\delta-1}}$-order of $\lambda_p$. These $p$ charge
assignments may be identical or different, but come with an order. We
require as well the condition that
$\sum_{h=1}^pq_p^{(h)}[0_p]=0$, where the weight $0_p$ denotes the  
$\widehat{{\rm su}}(n/p)_{k/p}$ vacuum $0_p=(k/p;0,\ldots,0)$.
(This condition will guarantee that we end up with resolved charges
for SU$(n)/\Zop_d$.) Then we get a resolved charge 
assignment $q$ for SU$(n)/\Zop_{p^\delta}$ level $k$, by defining
$q[\lambda]\equiv 0$ if $\lambda$ is not fixed by a non-trivial 
power of $J^{n/p^\delta}$, and defining 
$q[\phi,(i-1)p+h]\equiv q^{(h)}_p [\phi^p,i]$ for any fixed point
$\phi$. (For each fixed point $\phi$ of $J^{n/p^{\delta}}$, 
$\phi^p=\bar{\phi}$ means to truncate it after $n/p$ components.)

Using this identification, the resolved charges for
SU$(n)/\Zop_{p^\delta}$ level $k$ can be built up recursively 
from the charges of SU$(n/p^i)/\Zop_{p^{\delta-i}}$ level $k/p^i$.
Note that in order for this to work we need that 
$M(\widehat{{\rm su}}(n)_k)=M(\widehat{{\rm su}}(n/p)_{k/p})$, as 
is readily verified from (\ref{M_A}).
\medskip

The relation of {\it unresolved} charges for SU$(n)/\Zop_{d}$ to those
for SU$(n)/\Zop_{p^\delta}$ is as for the resolved ones, except that 
for each prime $p$ dividing $M$ but not $d$, we now want
$q[\phi,i]=(o(\phi))^{-1}{\rm dim}(\phi)$ (mod $p^\mu$) 
($o(\phi)$ will be coprime to $p$ and hence also invertible mod $p^\ell$).
At present we know of no direct relation between unresolved charges of
SU$(n)/\Zop_{p^\delta}$ and those of SU$(n/p)/\Zop_{p^{\delta-1}}$, 
although fixed point factorisation is surely involved, and
identifying those unresolved charges is the only remaining issue
here. 

We conjecture that an order $M$ untwisted charge can always be found.
If true, this would imply that the ${\cal D}$ charge group is given by
(\ref{noddres}). Infinite classes where this is known to hold 
({\it e.g.}\ $M$ coprime to $d$) are given in
\cite{Gaberdiel:2004yn}. Another, generalising the example
SU$(6)/\Zop_3$ given in \cite{Gaberdiel:2004yn}, is 
SU$(mp)/\Zop_p$ where $1<m<p$. It suffices to consider there levels
$k$ such that $p^2$ divides $k+mp$ (otherwise $M$ would be coprime to
$p$). We claim that $p$ will divide any fixed point dimension
dim$(\phi)$ here, and hence that $q[\phi,i]={\rm dim}(\phi)/p$
works. This follows directly from Weyl's dimension formula
\begin{eqnarray}
{\rm dim}(\phi)& = & 
{({k+mp\over p})^{mp(p-1)/2}\left(\prod_{\ell=1}^{p-1} \ell^{m(p-1)}
\prod_{1\le i<j\le m}(\ell^2({k+mp\over p})^2-(\phi(j)-\phi(i))^2)\right)
\over 
p^{pm(m-1)/2}\left(\prod_{\ell=1}^{m-1} \ell^{p(m-1)}
\prod_{1\le i<j\le p}(\ell^2p^2-(j-i)^2)\right) } \nonumber \\
& & \quad \qquad \times {
\left(\prod_{1\le i<j\le m}(\phi(j)-\phi(i))^p\right)\over
\left(\prod_{1\le i<j\le p}(j-i)^m\right)} \ , 
\end{eqnarray}
where $\phi(i)\equiv\sum_{l=1}^i\phi_l$. Counting the occurrences of
$p$, we get at least $mp(p-1)/2$ in the numerator and exactly
$pm(m-1)/2$ in the denominator. Thus for SU$(mp)/\Zop_p$ when $1<m<p$,
at any level $k$, the ${\cal D}$ charge group is 
$\Zop_M\oplus (p-1)\cdot \Zop_p$ or $\Zop_M$, depending on whether or 
not $p^2$ divides $k+mp$, in perfect agreement with (\ref{noddres}).

\section{Comparing charge groups of ${\cal D}$ and ${\cal D}^*$}

As we have mentioned before, the charge groups for ${\cal A}$ and
${\cal A}^*$ are known to agree
\cite{Gaberdiel:2003kv,GV,Fredenhagen:2005cj} in all cases. We have 
furthermore seen evidence in the above that the same is true for the  
charge groups for ${\cal D}$ and ${\cal D}^*$ --- indeed our two
conjectures imply that they will always agree in the non-pathological
cases, and be given by (\ref{noddres}). From the point of view of this
paper this agreement is quite remarkable, considering that the 
analysis for ${\cal D}$ and ${\cal D}^*$ appears to be 
very different.

On the other hand, we have also found a {\it pathological} example in
which the ${\cal D}$ and ${\cal D}^*$ charge groups do not
coincide.\footnote{MRG thanks Pedro Schwaller for helping him check
this carefully.} Indeed, for 
SU$(4)/\Zop_4$ at $k=4$, the ${\cal D}$ charge group equals 
$\Zop_4\oplus 2\cdot\Zop_2$, while the ${\cal D}^*$ charge group is
$2\cdot\Zop_4$. Again this seems to show that the pathological case is
structurally rather different. It would be interesting to see whether
the other modular invariant \cite{Fredenhagen:2004xp} that can be
defined in this case behaves better.

Nevertheless the striking agreement for the non-pathological cases
suggests that there is a more conceptual way in which this can be
understood. In the following we describe some preliminary steps
towards this goal.

\subsection{The intertwiner}

It is more natural to think about the correspondence between the two
charge groups in the context where we consider, for the modular
invariant (\ref{modular}) say, the untwisted and the twisted
D-branes. The open strings between the untwisted D-branes are then
characterised by the NIM-rep ${\cal D}$, while those
between the twisted D-branes are described by 
${\cal D}^*$. If we consider both sets of branes simultaneously, we
can therefore combine the two NIM-reps into
\begin{equation}\label{5.M}
{\cal N}^{{\rm full}}_\la=
\left(\begin{matrix}
{\cal N}({\cal D})_\la& 0\\ 0&{\cal N}({\cal D}^*)_\la
\end{matrix}\right)\ .
\end{equation}
However, we can now also consider the open strings between an
untwisted and a twisted D-brane. They will transform in a twisted
representation of the affine algebra, and thus we have in addition 
the off-diagonal matrix \cite{Gaberdiel:2002qa}
\begin{equation}\label{5.e}
{\cal N}^{{\rm full}}_{\hat{a}}=
\left(\begin{matrix} 0 & \rho_{\hat{a}} \\
\rho^t_{\hat{a}} & 0 
\end{matrix}\right)\ .
\end{equation}
Given the explicit formulae for the $\psi$-matrix of the boundary
states, it is in principle straightforward to calculate these matrix
elements. However, it would be useful to have explicit formulae for
these matrix elements in terms of suitable fusion rule
coefficients. For example, if $n$ is odd and gcd$(d,k)=1$, there are
no fixed points, and one simply finds that 
\begin{equation}
\rho_{\hat{a},[\nu]}{}^{(\hat{b},i)} 
= \N({\cal A}^*)_{\nu\,\hat{a}}{}^{\hat{b}} \ ,
\end{equation}
{\it i.e.}\ that $\rho$ agrees with the twisted fusion rules. It
is not difficult to check that this definition is well-defined 
({\it i.e.}\ independent of which representative $\nu$ in the orbit
$[\nu]$ is chosen). Furthermore, the resulting full 
NIM-rep forms indeed a representation of the full fusion ring that
includes untwisted and twisted representations
\cite{Gaberdiel:2002qa} (see also \cite{IT}). This last property
implies, in particular, that $\rho$ defines an intertwiner between the
two NIM-reps, {\it i.e.}  
\begin{equation}\label{nimin}
\sum_{(\hat{b},r)} \rho_{\hat{a},([\nu],i)}{}^{(\hat{b},r)}  \;\; 
{\cal N}({\cal D}^*)_{\lambda, (\hat{b},r)}{}^{(\hat{c},s)} 
=
\sum_{([\mu],j)} {\cal N}({\cal D})_{\lambda, ([\nu],i)}{}^{([\mu],j)}
\;\;
\rho_{\hat{a},([\mu],j)}{}^{(\hat{c},s)}  \ .
\end{equation}
Here, as in the following, we are labelling the boundary states
of the ${\cal D}^*$ NIM-rep by $(\hat{b},r)$; this is, as we have
seen, appropriate as long as the theory is not pathological. The
boundary states of the ${\cal D}$ NIM-rep are labelled by 
$([\nu],i)$. 
\medskip

This off-diagonal matrix $\rho$ defines therefore a map from the
untwisted to the twisted D-branes that intertwines the NIM-rep
action. Furthermore, there is a canonical
`smallest' twisted representation, namely the representation $\hat{0}$
that has a one-dimensional highest-weight space. It is therefore
natural to guess that the off-diagonal matrix associated to $\hat{0}$
maps the minimal charged D-brane of ${\cal D}$ to the minimal charged
D-brane of ${\cal D}^*$, and that this can `explain' the equivalence
of the charge groups. In fact, this is how the situation works for 
the case of the untwisted and twisted NIM-reps ${\cal A}$ and 
${\cal A}^*$ of the simply connected group. In that case $\rho$ can
naturally be identified with the twisted fusion rules themselves, and
the minimal solutions are indeed related as  
\begin{equation}\label{inter}
{\rm dim}(\lambda) = \sum_{\hat{a}} 
\N({\cal A}^*)_{\lambda\,\hat{0}}{}^{\hat{a}}\;
{\rm dim}(\hat{a}) \quad
{\rm (mod}\ M)\ .
\end{equation}
One may hope to be able to use this idea to {\it prove} the a priori
non-trivial fact that the two charge groups of ${\cal A}$ and 
${\cal A}^*$ must be the same, but we have not yet succeeded in doing
so. The main problem is that we do not understand how to invert this
relation, {\it i.e.}\ how to express the charges of the twisted branes
dim$(\hat{a})$ in terms of those of the the untwisted branes
dim$(\lambda)$. 
\smallskip

The analogue of (\ref{inter}) can always be defined: suppose that we
have a solution $q_{{\cal D}^*}(\hat{a},i)$ of the ${\cal D}^*$ charge
equations
mod some $M$.  Then we can define a solution of the ${\cal D}$ charge
equations by   
\begin{equation}\label{ansatz}
q_{{\cal D}}([\nu],l) \equiv \sum_{(\hat{a},i)} \, 
\rho_{\hat{0},([\nu],l)}{}^{(\hat{a},i)} \, q_{{\cal D}^*}(\hat{a},i) \ .
\end{equation}
It is not difficult to see that the $q_{{\cal D}}([\nu],j)$ then satisfy
the ${\cal D}$ charge equations mod the same $M$. Indeed, we have 
\begin{eqnarray}
\dim(\lambda) \, q_{{\cal D}}([\nu],l) 
& = & \sum_{(\hat{a},i)} \, 
\rho_{\hat{0},([\nu],l)}{}^{(\hat{a},i)} \, \dim(\lambda)\, 
q_{{\cal D}^*}(\hat{a},i) \nonumber \\ 
& = & \sum_{(\hat{a},i)} \, \sum_{(\hat{b},j)}
\rho_{\hat{0},([\nu],l)}{}^{(\hat{a},i)} \, 
{\cal N}({\cal D}^*)_{\lambda, (\hat{a},i)}{}^{(\hat{b},j)}  \, 
q_{{\cal D}^*}(\hat{b},j) \quad {\rm (mod}\ M) \nonumber \\
& = & \sum_{([\mu],m)} \, \sum_{(\hat{b},j)}
\rho_{\hat{0},([\mu],m)}{}^{(\hat{b},j)} \, 
{\cal N}({\cal D})_{\lambda, ([\nu],l)}{}^{([\mu,m])}  \, 
q_{{\cal D}^*}(\hat{b},j) \nonumber \\
& = & \sum_{([\mu],m)} {\cal N}({\cal D})_{\lambda,
([\nu],l)}{}^{([\mu,m])}  \, q_{{\cal D}}([\mu],m) \ ,
\end{eqnarray}
where we have used the NIM-rep property (\ref{nimin}) in the third
line.

\subsection{Symmetries and resolutions}

While (\ref{ansatz}) associates to any solution of the ${\cal D}^*$
charge equations a solution of the ${\cal D}$ charge equations, it is
a priori not clear whether all different solutions of ${\cal D}$ can
be obtained in  this manner. In fact, it is not difficult to find a
counterexample to this: already for SU$(3)/\Zop_3$ at level $k=3$ the
above map does not produce all different solutions of the ${\cal D}$
charge equations. However, it is also not difficult to understand the
reason for this: at $k=3$ the SU$(3)/\Zop_3$ has additional
symmetries, and these can be used to `resolve' the intertwiner
$\rho_{\hat{0}}$ further. To understand how this can be done, we 
observe that one can associate 
conserved charges to the boundary states.\footnote{These charges
should {\it not} be confused with the D-brane charges we have
discussed before!} For the case of the ${\cal D}^*$ boundary states,
the construction is easy: we simply define $Q(\hat{b},r)=r$. This
gives indeed a conserved charge since  
\begin{equation}
{\cal N}({\cal D}^*)_{\lambda,(\hat{b},r)}{}^{(\hat{c},s)} \ne 0 \
\Longrightarrow \  t(\lambda) + Q(\hat{b},r) = Q(\hat{c},s) \quad
{\rm (mod}\ d) \ . 
\end{equation}
What are the possible conserved charges for the ${\cal D}$ boundary
states? In general, the charge $Q([\nu],i)$ cannot depend on $i$ since
the ${\cal D}$ NIM-rep is independent of $i$ if $\nu$ is not a fixed
point. The only conserved charges of the fusion ring of su$(n)$ are
multiples of the $n$-ality, but for the quotient theory in question we
need them to be independent of the representatives in the $J^{d'}$
orbits. One easily checks that  
\begin{equation}
t\left(J^{d'} \nu\right)  = d' k + t(\nu) \ .
\end{equation} 
This means that we can define $Q([\nu],i)=t(\nu)$, but that the
resulting charge is only defined mod $R'={\rm gcd}(n,nk/d)$
(since $n$-ality is only defined mod $n$). Since the charges for the
${\cal D}^*$ boundary states are only defined mod $d$, we have
combined charges defined mod $R={\rm gcd}(d,R') = {\rm gcd}(d,d'k)$.  

The idea is now that we can `resolve' each $\rho$-intertwiner (and in
particular the one associated to $\hat{0}$) into $R$ intertwiners. 
Explicitly we define
\begin{equation}
\rho_{[\hat{0},x],([\nu],m)}{}^{(\hat{a},i)} \equiv
\rho_{\hat{0},([\nu],m)}{}^{(\hat{a},i)} \, 
\delta^{(R)}\left(Q(\hat{a},i) - Q([\nu],m)-x\right) \ . 
\end{equation}
By construction, we have 
\begin{equation}
\rho_{\hat{0}} = \sum_{x=1}^{R} \rho_{[\hat{0},x]} \ .
\end{equation}
Thus the $\rho_{[\hat{0},x]}$ define a `resolution' of the original
intertwiner $\rho_{\hat{0}}$. Each $\rho_{[\hat{0},x]}$ is in fact
separately an intertwiner: this follows directly from the fact that 
$\rho_{\hat{0}}$ is, together with the property of the two charges 
$Q(\hat{a},i)$ and $Q([\nu],m)$ to be conserved.

In particular, each resolved intertwiner $\rho_{[\hat{0},x]}$ can
therefore also play the role of $\rho_{\hat{0}}$ in (\ref{ansatz}).  
In fact, for SU$(3)/\Zop_3$ at $k=3$ we have $R=3$, and we can
therefore resolve $\rho_{\hat{0}}$ into $3$ intertwiners. One then
easily checks that these resolved intertwiners now {\it do} account
for all the different charge solutions for the ${\cal D}$ NIM-rep in
terms of those of the ${\cal D}^*$ NIM-rep. We suspect that this will
always hold, {\it i.e.}\ that (\ref{ansatz}) with $\rho_{\hat{0}}$
replaced by the resolved intertwiners $\rho_{[\hat{0},x]}$ always
accounts for all charge solutions of ${\cal D}$ in terms of those of
${\cal D}^*$. However, we have so far not been able to prove this for
the general (non-pathological) case.

\section{Conclusions}

In this paper we have studied the twisted branes of the WZW model
corresponding to the non-simply connected group manifold
SU$(n)/\Zop_d$. In particular, we have found a very explicit formula
for the multiplicities (NIM-rep) with which the different affine
representations appear in the relative open strings between 
these branes. At least in the non-pathological case ($n(n+1)/d$ even)
this NIM-rep is remarkably simple: if $n$ is odd it is given by
(\ref{D*odd}), while for $n$ even we have instead (\ref{3.3}) subject
to the constraint (\ref{nevrest}). 

Given these simple formulae, we have calculated the charge group that
is generated by these twisted D-branes. For $n$ odd we have proven
that the result is given by (\ref{noddres}). We have also given some 
convincing evidence that the same result holds for even $n$.  
\smallskip

We have also made progress towards the description of the NIM-rep for
the untwisted branes. In particular, using fixed point factorisation
techniques, we have found simple formulae for the NIM-rep coefficients
for the fundamental weights (\ref{4.8a}) and (\ref{4.8b}). Using these
explicit expressions we have managed to perform a more careful
analysis of the corresponding charge group than was possible in 
\cite{Gaberdiel:2004yn}. While we have not succeeded in determining it
in general, we have analysed it for some examples (in particular
SU$(9)/\Zop_9$ at level $k=18$, as well as an infinite class of the
form SU$(n)/\Zop_p$ for all $k$), and we have given good evidence that
it is again given by (\ref{noddres}). 

Our results thus provide convincing support for the assertion that the
charge groups for the untwisted and twisted D-branes agree also for
the non-simply connected WZW models SU$(n)/\Zop_d$, provided that the
theory is not pathological, {\it i.e.}\ that $n(n+1)/d$ is even.  
If we turn the argument around and {\it assume} that the two charge
groups are equal in the non-pathological cases, it would be easy to
prove that both are given by (\ref{noddres}). In fact, the only gap in
our argument for the ${\cal D}$-charge group concerns the unresolved 
solution that gives rise to the summand $\Zop_{M}$; the existence of
this solution is immediate in the ${\cal D}^*$ case. Conversely, 
the gap in the argument for ${\cal D}^*$ concerns the different
solutions with $q(\widehat{0},i)\ne q(\widehat{0},j)$ that correspond
to the resolved solutions in the ${\cal D}$ case. Given our improved 
understanding of the ${\cal D}$-NIM-rep these are now under good
control. It would therefore be very interesting if one could establish
the equivalence of these two charge groups abstractly; first steps in
this direction were described in section~5. 

It would also be very interesting to calculate the relevant 
K-theory groups  using a geometrical approach, 
and compare the results
with the predictions of our conformal field theory analysis. Since the
charge groups that were derived above have a very rich structure, this
would be a very convincing consistency check of the whole approach.

\section*{Acknowledgements}

\noindent We thank Stefan Fredenhagen for useful
conversations. This research was done while TG was visiting Hamburg
University as a Humboldt Fellow; he also thanks ETH for hospitality during
a fruitful visit. His
research is partly supported by NSERC. The research of MRG is
supported in part by  the Swiss National Science Foundation and the
Marie Curie network `Constituents, Fundamental Forces and Symmetries
of the Universe' (MRTN-CT-2004-005104).

\appendix

\section{Proof of Proposition 2}

In this appendix we give the proof of Proposition 2(a). First note that
$\bar{D}_e$ must divide $n^e$. To see this, expand the tensor
product $\Lambda_1\otimes\cdots\otimes \Lambda_1$ out into a sum of 
irreducible $\la$, and hence write dim$(\Lambda_1)^e=n^e$ as a sum
(with multiplicities) of certain dim$(\la)$, where each 
$t(\la)= e$ (mod $n$). Since $\bar{D}_e$ divides each   
such dim$(\la)$, it must divide their sum.

Let $p^\nu\|n$. Then repeatedly applying Proposition 1(b), we get for any
$j$ that 
\begin{equation}\label{2.C}
{\rm dim}(\Lambda_{p^j\ell})= \left({n/p^j\atop \ell}\right)\quad
 ({\rm mod}\ p^{\nu-j+1})\ .
\end{equation}
Thus if $p$ does not divide $d$, then 
$({\rm dim}(\Lambda_{p^\nu}))^e= (n/p^\nu)^e$ (mod $p$) will be
coprime to $p$, but as in the previous paragraph must be divisible by
$\bar{D}_e$. This tells us that $\bar{D}_e$  must in fact
divide $d^\infty$. 

Now suppose $p$ divides $d$. Taking the gcd of eq.(\ref{2.C}), over
all $0<\ell<n/p^j$  coprime to $p$, we get $p^{\nu-j}$ (by Proposition 1(a)
applied to su($n/p^j$)).  So $p^{\nu-j}$ divides each
dim$(\Lambda_{p^j\ell})$, when $\ell$ is coprime to $p$. 

Now, consider any weight $\la$ with gcd$(t(\la),\bar{D})=e$. The
su$(n)$-character ch$_\la$ will be a polynomial, with coefficients in
$\Zop$, in the fundamental weights ch$_{\Lambda_\ell}$, so dim$(\la)$
will equal that same polynomial, evaluated at dim$(\Lambda_\ell)$. This
polynomial will be homogeneous (in the obvious weighted sense) of
degree $t(\la)$ (mod $n$). Let $p^\epsilon\|e$, for some prime $p$. We
first want to show that $p^{\nu-\epsilon}$ divides each term in that
polynomial, {\it i.e.}\ that it divides each product $\prod_i{\rm dim}
(\Lambda_{\ell_i})$ of  total weighted degree $t\equiv \sum_i \ell_i$ 
satisfying gcd$(t,\bar{D})=e$.

But this is clear. The given term must contain some 
dim$(\Lambda_{\ell p^h})$, where $\ell$ is coprime to $p$ and 
$0\le h\le \epsilon$. So $p^{\nu-h}$ (and hence $p^{\nu-\epsilon}$)
will divide dim$(\Lambda_{\ell p^h})$ and hence the whole 
term.\footnote{We are being a little sloppy here: when $\epsilon=\nu$
it is possible for all $h$ to exceed $\epsilon$, but in that boundary
case we only have to prove that $p^{\nu-\epsilon}=1$ divides the
dimensions, which is trivial.} Repeating for all $p$, we find that   
$\bar{D}/e=\prod_{p|d}p^{\nu-\epsilon}$ will divide
$\bar{D}_e$. 

All that remains is to show the other direction, {\it i.e.}\ for each
prime $p$ to  find a weight $\la$ with gcd$(t(\la),\bar{D})=e$,
and with $p^{\nu- \epsilon}\|{\rm dim}(\la)$. To do this, consider the
tensor product $\Lambda_{p^\epsilon}\otimes\Lambda_{p^{\nu}t}$, where
we take $0\le t <n/ p^\nu$ so that $p^\epsilon+p^\nu t= e$ (mod
$n$). From eq.(\ref{2.C}), $p^{\nu-\epsilon}$ exactly divides
dim$(\Lambda_{p^\epsilon})$, while $p$ is coprime to 
$\Lambda_{p^\nu t}$, and so $p^{\nu-\epsilon}$ exactly  
divides that product of dimensions. Expanding that product out into a
sum of dimensions, $p^{\nu-\epsilon}$ will divide each dimension
separately in that sum, and so it must divide exactly at least one
dimension in that sum. That is the desired dim$(\la)$. This argument
breaks down when $p^\epsilon=n$, but in this case
$\la=\Lambda_1+\Lambda_{n-1}$, with dimension $n^2-1$ coprime to $p$,
works. 
\hspace*{\fill} {\bf QED}.

\section{Fixed point factorisation and NIM-reps}

Consider any non-pathological SU$(n)/\Zop_d$ at level $k$ 
($n$ can be even or odd). Let $f={\rm gcd}(d,k)$ and write $d'=n/d$ as
usual.   

Let $[\varphi,i]$, $[\varphi',i']$ be boundary states with orders 
$o(\varphi)=g$, $o(\varphi')=g'$, respectively. Let $(\psi,j)$ be an
exponent with order $o(\psi)=h$. We will let superscript $(\delta)$
denote quantities associated with $\widehat{\rm su}(n/\delta)$ at
level $k/\delta$. We let $0$ and $J$ denote the vacuum and simple
current generator in any $\widehat{\rm su}(n/\delta)$ level
$k/\delta$.

Assume we are in {\bf Case A} (this includes all non-pathological, but 
also some pathological, cases). Then the formula of 
\cite{FSS1,Birke:1999ik} is  
\begin{equation}\label{A.4}
\psi_{[\varphi,i],(\psi,j)}={\sqrt{d}\over g\sqrt{h}}
\sum_{\delta|{\rm gcd}
(g,h)}\xi_\delta\,s(\delta,i-j)\,
S^{(\delta)}_{\varphi^\delta\psi^\delta} \ ,
\end{equation}
where $\xi_\delta$ is some irrelevant root of unity depending on  
$\delta,n,k,d$, and 
\begin{equation}\label{A.5a}
s(a,b)=\sum_{\ell\in\Zop_a^\times}e^{2\pi\i\ell b/a}
\end{equation}
(that is, the sum is over all $1\le \ell\le a$ coprime to $a$). 
In fact, this quantity $s(a,b)$ is called a Ramanujan sum, and is easily 
seen to equal
\begin{equation}\label{A.5b}
s(a,b)=\sum_{d|{\rm gcd}(a,b)}\mu(n/d)\,d \ , 
\end{equation}
where $\mu(c)$ is the {\it M\"obius function}, which equals $0$ unless
$c$ is the product of $s\ge 0$ distinct primes, in which case 
$\mu(c)=(-1)^s$. We will need one other property of   $s(a,b)$:
provided $a,b$ divide $h$, 
\begin{equation}\label{A.5c}
\sum_{j=1}^h s(a,i-j)\,s(b, j-i')=h \,s(a,i-i')\,\delta_{a,b} \ .
\end{equation}
This identity follows directly from (\ref{A.5a}) and the calculation
\begin{align}
\sum_{j=1}^h\sum_{\ell\in\Zop_a^\times}
\sum_{m\in\Zop_b^\times} & 
e^{2\pi\i\ell(i-j)/a}e^{2\pi\i m(j-i')/b}\cr 
=&\, \sum_{\ell\in\Zop_a^\times}
\sum_{m\in\Zop_b^\times}\exp[2\pi\i({\ell i\over a}-{mi'\over b})]
\sum_{j=1}^h
\exp\left[2\pi\i {j\over h}({mh\over b}-{\ell h\over a})\right]\cr 
=&\,\sum_{\ell\in\Zop_a^\times}
\sum_{m\in\Zop_b^\times}\exp[2\pi\i({\ell i\over a}-{mi'\over b})]
\delta_{\ell,m}\delta_{a,b}=h \,s(a,i-i')\,\delta_{a,b} \ . 
\end{align}
Plugging everything into (\ref{nimpsi}) gives a mess, even at the
fundamental weights:  
\begin{align}\label{A.6}
\N_{\Lambda_m, [\varphi,i]}{}^{[\varphi',i']}
={d\over gg'}\sum_{h|f}
\sum_{t(\psi)=_d 0\atop o(\psi)=h}{1\over h}\sum_{j=1}^h&\left(
\sum_{\delta|{\rm gcd}(g,h)}
\xi_\delta\,s(\delta,i-j)\,S^{(\delta)}_{\varphi^\delta\psi^\delta}
\right){S_{\Lambda_m\psi}\over S_{0\psi}}&\cr
&\quad \left(\sum_{\delta'|{\rm gcd}(g',h)}
\xi_{\delta'}^*\,s(\delta',j-i')\,S^{(\delta')*}_{\varphi'{}^{\delta'}
\psi^{\delta'}}\right) \ .
\end{align}
The key observation is that, because of (\ref{A.5c}), the crossterms
in (\ref{A.6}) (that is the terms with
$\delta\ne\delta'$) vanish, and what we obtain is 
\begin{equation}
\N_{\Lambda_m, [\varphi,i]}{}^{[\varphi',i']}={d\over gg'}\sum_{h|f}
\sum_{t(\psi)=_d 0\atop o(\psi)=h}\sum_{\delta|{\rm gcd}(g,h,g',m)}
s(\delta,i-i')\,S^{(\delta)}_{\varphi^\delta \psi^\delta}{S^{(\delta)}_{
\Lambda_{m/\delta}\psi^\delta}\over S^{(\delta)}_{0^\delta\psi^\delta}}
S^{(\delta)}_{\varphi'{}^\delta \psi^\delta} \ ,
\end{equation}
where we have used fixed point factorisation.
The restriction to $t(\psi)= 0$ (mod $d$) is easily obtained by a
sum over simple currents: ${1\over d}\sum_{j=1}^d$. However, we can 
write ${d\over g''}=a{d\over g}+b{d\over g'}$, where 
$g''={\rm lcm}(g,g')$, so by (\ref{1.1a}) it suffices 
to take the sum ${g''\over d}\sum_{j=1}^{d/g''}$. We obtain
\begin{equation}\label{A.7a}
\N_{\Lambda_m, [\varphi,i]}{}^{[\varphi',i']}={g''\over gg'}
\sum_{\delta|{\rm gcd}(g,g',m)}s(\delta,i-i')\,\sum_{j=1}^{d/g''}
N^{(\delta)}_{J^{d'j}
\Lambda_{m/\delta}\,\varphi^\delta}{}^{\varphi'{}^\delta}\ .
\end{equation}
Again, $N^{(\delta)}$ denotes fusion coefficients for
$\widehat{\rm su} (n/\delta)$ level $k/\delta$. By substituting in
(\ref{A.5b}) and rearranging the sum, we can rewrite (\ref{A.7a}) in a
more friendly form: 
\begin{equation}\label{A.7b}
\N_{\Lambda_m, [\varphi,i]}{}^{[\varphi',i']}=
{g''\over gg'}\sum_{j=1}^{d/g''} \ 
\sum_{a|\Delta}a\delta^{(a)}_{ii'} 
\sum_{\delta, \ a|\delta|\Delta}\mu (\delta/a)\,
N^{(\delta)}_{J^{d'j}\Lambda_{m/\delta}\,
\varphi^\delta}{}^{\varphi'{}^\delta} \ ,
\end{equation}
where $\Delta={\rm gcd}(g,g')$, and where $\delta^{(x)}_{ij}$ equals
$0$ unless $x$ divides $i-j$, when it equals 1. The second sum is over
all multiples $\delta$ of $a$, which divide $\Delta$. If some $\delta$
does not divide $m$ then we interpret the entire right-side of
(\ref{A.7b}) as equalling $0$. That the coefficient
$\N_{\Lambda_m,[\varphi,i]}^{[\varphi',i']}$ must vanish 
if $\Delta$ (or any other $\delta$) does not divide $m$, is clear from
(\ref{nimsum}) and positivity. 

A special case of (\ref{A.7b}) will be very useful. If $\Delta$ equals a
prime $p$, and $d=g''$ and $p$ divides $m$, then (\ref{A.7b})
simplifies to  
\begin{equation}
\N_{\Lambda_m, [\varphi,i]}{}^{[\varphi',i']}=
{1\over p}\left(N_{\Lambda_{m}\,
\varphi}{}^{\varphi'}-N^{(p)}_{\Lambda_{m/p}\,
\varphi^p}{}^{\varphi'{}^p}
+p\delta^{(p)}_{ii'}N^{(p)}_{\Lambda_{m/p}\,
\varphi^p}{}^{\varphi'{}^p}\right) \ .
\end{equation}
But these coefficients must be integral, and thus 
$N_{\Lambda_{m}\,\varphi}^{\varphi'}$ and
$N^{(p)}_{\Lambda_{m/p}\,\varphi^p}{}^{\varphi'{}^p}$ 
should be congruent (mod $p$). But by the Pieri formula, 
tensor product coefficients (hence fusion coefficients) involving
fundamental weights will always be 0 or 1. Thus we obtain
\begin{equation}\label{A.7c}
N_{\Lambda_{m}\,\varphi}{}^{\varphi'}=
N^{(p)}_{\Lambda_{m/p}\,\varphi^p}{}^{\varphi'{}^p} \ ,
\end{equation}
which must hold for any multiples $n,k,m$ of $p$, and weights  
$\varphi,\varphi'$ fixed by $J^{n/p}$. Using this, (\ref{A.7b})
simplifies when $\Delta=p$, but as we will see shortly much more is
true. 

Even  formula (\ref{A.7b}) is surprisingly simple. But we can reduce
it much more, by induction on $d$. Suppose for all $n,k$, we get the
following formula --- the main result of this appendix --- valid 
for any possible $d<D$ (of course $d$ must divide $n$):
\begin{eqnarray}
\N_{\Lambda_m,[\varphi,i]}{}^{[\varphi',i']} & =&\,
\sum_{j=1}^{d/g''}
\delta_{ii'}^{(\Delta)}N^{(\Delta)}_{J^{d'j}\Lambda_{m/\Delta}
\varphi^\Delta}{}^{\varphi'{}^\Delta} \label{A.8a} \\
\N_{J0,[\varphi,i]}{}^{[\varphi',i']}& = &\,
\delta_{[\varphi'],[J\varphi]}\delta_{i,i'} \ ,
\label{A.8b}
\end{eqnarray}
where again we interpret the right-side of (\ref{A.8a},\ref{A.8b}) as 
vanishing if $\Delta$ does not divide $m$. We want to show that the
induction  hypothesis (\ref{A.8a}) will also hold for $d=D$. This
would then imply that (\ref{A.8a}) always holds.

Certainly the induction hypothesis holds for $D=2$. In fact from results
in \cite{Gaberdiel:2004yn} (or using (\ref{A.7c})) in (\ref{A.7b})) we know it
holds for $D=4$ as well. So  take $d=D$, and any multiple 
$n$ of $d$. We may assume that $\Delta={\rm gcd}(g,g')$ divides $m$.
Consider any prime  $p$ dividing $\Delta$ --- say it divides $\Delta$
exactly $\alpha$ times. Write $n_0=n/p^\alpha$, $d_0=d/p^\alpha$, $k_0=
k/p^\alpha$, $\Delta_0=\Delta/p^\alpha$. Note that $\varphi^{p^\alpha}$ and
$\varphi'{}^{p^\alpha}$ have orders $g_0=g/p^\alpha$ and $g_0'=g'/p^\alpha$.
We want to reduce (\ref{A.8a},\ref{A.8b}) for $n,k,d$,
to (\ref{A.8a},\ref{A.8b}) for $n_0,k_0,d_0$.

\noindent Assume first that $\alpha\ge 2$. Then by virtue of the
M\"obius function, we can write
\begin{eqnarray} \label{A.9a}
\sum_{a|\Delta}a\delta^{(a)}_{ii'} \sum_{\delta, \ a|\delta|
\Delta}\mu(\delta/a)\,
N^{(\delta)}_{\la(\delta)\,\varphi^\delta}{}^{\varphi'{}^\delta}
& = & \sum_{b=0}^{\alpha}p^b\delta^{(p^b)}_{ii'}\sum_{a_0|\Delta_0}a_0
\delta^{(a_0)}_{ii'} \sum_{\delta_0, \ a_0|\delta_0|\Delta_0}
\mu(\delta_0/a_0) \\
&& \left(
N^{(p^b\delta_0)}_{\la(p^b\delta_0)\,\varphi^{p^b\delta_0}}
{}^{\varphi'{}^{p^b
\delta_0}}-N^{(p^{b+1}\delta_0)}_{\la(p^{b+1}\delta_0)\,
\varphi^{p^{b+1}\delta_0}}{}^{\varphi'{}^{p^{b+1}\delta_0}}\right) \ ,
\nonumber 
\end{eqnarray}
where we drop all terms with $p^{\alpha+1}$. The weight we call
$\la(\delta)$ here is $J^{d'j}\Lambda_{m/\delta}$, but it could be
anything. If instead $\alpha=1$, then again (\ref{A.9a}) must hold. 

The term in the brackets of (\ref{A.9a}) equals $0$, by virtue of
(\ref{A.7c}). Thus everything on the right-side of (\ref{A.9a})
vanishes, except for the $b=\alpha$ terms, and we get
\begin{align}\label{A.9b}
\sum_{a|\Delta}a\delta^{(a)}_{ii'} & \sum_{\delta,\ a|\delta|\Delta} 
\mu(\delta/a)\,
N^{(\delta)}_{\la(\delta)\,\varphi^\delta}{}^{\varphi'{}^\delta} 
\cr & 
=p^\alpha\delta^{(p^\alpha)}_{ii'}\sum_{a_0|\Delta_0}a_0
\delta^{(a_0)}_{ii'} 
\sum_{\delta_0, \ a_0|\delta_0|\Delta_0}
\mu(\delta_0/a_0)\,
N^{(p^\alpha\delta_0)}_{\la(p^\alpha\delta_0)\,\varphi^{p^\alpha\delta_0}}
{}^{\varphi'{}^{p^\alpha \delta_0}} \ . 
\end{align}
Substituting this into (\ref{A.7b}), we obtain
\begin{align}
\N_{\Lambda_m, [\varphi,i]}{}^{[\varphi',i']}=&\,{g''_0\over p^\alpha 
g_0g'_0}
\sum_{j=1}^{d_0/g''_0}p^\alpha\delta^{(p^\alpha)}_{ii'}
\sum_{a_0|\Delta_0}a_0
\delta^{(a_0)}_{ii'} \sum_{\delta_0, \ a_0|\delta_0|\Delta_0}
\mu(\delta_0/a_0)\,
N^{(p^\alpha\delta_0)}_{J^{d'j}\Lambda_{{m\over p^\alpha}/\delta_0}\,
\varphi^{p^\alpha\delta_0}}{}^{\varphi'{}^{p^\alpha\delta_0}}\cr
=&\,
\N^{(p^\alpha)}_{\Lambda_{m\over p^\alpha}, [\varphi^{p^\alpha},i\ 
({\rm mod}\ p^\alpha)]}{}^{[\varphi'{}^{p^\alpha},i'\ 
({\rm mod}\ p^\alpha)]} \ .
\end{align}
By the induction hypothesis, this gives us (\ref{A.8a}), and we are
done.

\section{Proofs for ${\cal D}$ charges}

Consider any SU$(n)/\Zop_d$ level $k$ in Case A. Let $q[\lambda,i]$
be a twisted charge  solution, taken mod $M$, to 
${\cal D}(SU(n)/\Zop_d)$, {\it i.e.}\ $q[\lambda,i]=0$ unless
$\lambda$ is a fixed point of some non-trivial power of $J^{d'}$. Let
$\phi$ be such a fixed point, of order $o(\phi)$. Then considering the
charge equation for $0={\rm dim}(\phi)\,q[0]$, we find that full sums
are $0$: 
\begin{equation}
\sum_{h=1}^{o(\phi)} q[\phi,h]=0\quad({\rm mod}\ M)\ .
\end{equation}
Now, for any integer $m$ coprime to $o(\phi)$, the charge equation for
${\rm dim}(\Lambda_m)\,q[\phi,i]$ will be full sums, by (\ref{nimfus}),
and so will be 0 (mod $M$). By Proposition 2(b), the gcd of those
dim$(\Lambda_m)$ will be gcd$(o(\phi)^\infty,n)$. Since in addition
$Mq[\phi,i]=0$ (mod $M$), we have that $Dq[\phi,i]=0$ (mod $M$), where
$D={\rm gcd}(M,(o(\phi)^\infty,n)$. This tells us that the twisted
charge group is a subgroup of $\Zop_n^\infty$ and can be built up
out of the primes dividing $d$. (This argument requires $o(\Lambda_m)=1$.
We will run into problems here only if $k=2$, $m=n/2$, in which case
$o(\phi)=2$ and $m$ must be odd, but this would be Case B,
which is excluded.)

Now choose any prime $p$ dividing $d$, and put $p^\nu\|n$, $p^\delta\|d$.
Let $p^\ell$ exactly divide $o(\phi)$, {\it i.e.}\ $p^\ell$ is the
$J^{n/p^\nu}$-order of fixed point $\phi$. Write $p^\gamma={\rm gcd}(M,n,
p^\infty)$. We claim $i=j$ (mod $p^\ell$) implies $q[\phi,i]=q[\phi,j]$
(mod $p^\gamma$).

To see this, let $Q$ be the product of all primes $\ne p$ dividing $o(
\phi)$. By (\ref{4.8a}), we have
dim$(\Lambda_m)\,q[\phi,i]={\rm dim}(\Lambda_m)\,
q[\phi,j]$ (mod $M$) whenever $i=j$ (mod $p^\ell$). Now run over all $m$
coprime to $Q$ and apply Proposition 2(b).

We are now prepared to show that it suffices to analyse prime-power
orbifolds of SU$(n)$. In particular, given twisted charges $q[\phi,i]$
for ${\cal D}({\rm SU}(n)/\Zop_d)$, define $q_p[\phi,i]$ (for $1\le i
\le p$) to be $(o(\phi)/p^\ell)q[\phi,i]$. Then the $q_p$ will solve
the charge equations (mod $p^\gamma$) for ${\cal D}({\rm SU}(n)/\Zop_{
p^\delta})$. The converse is also true: given twisted charges $q_p$
for ${\cal D}({\rm SU}(n)/\Zop_{p^\delta})$, for each prime dividing
$d$, you can get a twisted solution $q$ for ${\cal D}({\rm SU}(n)/
\Zop_d)$, defined by the same formula. This establishes a bijection
between the twisted charge group of ${\cal D}({\rm SU}(n)/\Zop_d)$ and
the direct product over $p$ of the full ${\cal D}$-charge groups of 
SU$(n)/\Zop_{p^\delta}$. 
The proof is straightforward, using (\ref{4.8a}), although the book-keeping
is messy.

Similar arguments explain how the {\it untwisted} ${\cal D}$-charges of
SU$(n)/\Zop_d$ are related to the ${\cal D}$-charges for
${\rm SU}(n)/\Zop_{p^\delta}$. The relation between twisted
${\cal D}$-charges for ${\rm SU}(n)/\Zop_{p^\delta}$ and those of
${\rm SU}(n)/\Zop_{p^{\delta-1}}$, follows directly from (\ref{4.8a}).

%\bibliographystyle{JHEP-2}
%\bibliography{references}
%\providecommand{\href}[2]{#2}\begingroup\raggedright

\end{document}